\documentclass[a4paper,fleqn,usenatbib]{mnras}

\usepackage{caption}
\usepackage{newtxtext,newtxmath}
\usepackage[T1]{fontenc}
\usepackage{ae,aecompl}
\usepackage{graphicx}
\usepackage{amsmath}
\usepackage{amssymb}
\usepackage{multirow,array}
\usepackage{adjustbox}
\usepackage{color,soul}
\usepackage[dvipsnames]{xcolor}
\usepackage{float}

\renewcommand{\theenumi}{(\arabic{enumi})}

\title[Clustering in galaxy feature data]{Reproducible $k$-means clustering in galaxy feature data from the GAMA survey}

\author[Sebastian Turner et al.]{Sebastian Turner$^{1}$\thanks{E-mail: s.turner1@2012.ljmu.ac.uk}, Lee S. Kelvin$^{1}$, Ivan K. Baldry$^{1}$, Paulo J. Lisboa$^{2}$, Steven N.
\newauthor Longmore$^{1}$, Chris A. Collins$^{1}$, Benne W. Holwerda$^{3}$, Andrew M. Hopkins$^{4}$, and
\newauthor Jochen Liske$^{5}$
\\
$^{1}$Astrophysics Research Institute, Liverpool John Moores University, 146 Brownlow Hill, Liverpool, L3 5RF, UK\\
$^{2}$Department of Applied Mathematics, Liverpool John Moores University, Byrom Street, Liverpool, L3 3AF, UK\\
$^{3}$Department of Physics and Astronomy, 102 Natural Science Building, University of Louisville, Louisville KY 40292, USA\\
$^{4}$Australian Astronomical Observatory, 105 Delhi Rd, North Ryde, NSW 2113, Australia\\
$^{5}$Hamburger Sternwarte, Universit\"at Hamburg, Gojenbergsweg 112, 21029 Hamburg, Germany\\
}

\date{Accepted XXX. Received YYY; in original form ZZZ}

\pubyear{2018}

\hypersetup{draft}
\begin{document}
\label{firstpage}
\pagerange{\pageref{firstpage}--\pageref{lastpage}}
\maketitle

\begin{abstract}
A fundamental bimodality of galaxies in the local Universe is apparent in many of the features used to describe them. Multiple sub-populations exist within this framework, each representing galaxies following distinct evolutionary pathways. Accurately identifying and characterising these sub-populations requires that a large number of galaxy features be analysed simultaneously. Future galaxy surveys such as LSST and Euclid will yield data volumes for which traditional approaches to galaxy classification will become unfeasible. To address this, we apply a robust $k$-means unsupervised clustering method to feature data derived from a sample of 7338 local-Universe galaxies selected from the Galaxy And Mass Assembly (GAMA) survey. This allows us to partition our sample into $k$ clusters without the need for training on pre-labelled data, facilitating a full census of our high dimensionality feature space and guarding against stochastic effects. We find that the local galaxy population natively splits into $2$, $3$, $5$ and a maximum of $6$ sub-populations, with each corresponding to a distinct ongoing evolutionary mechanism. Notably, the impact of the local environment appears strongly linked with the evolution of low-mass ($M_{*} < 10^{10}$ M$_{\odot}$) galaxies, with more massive systems appearing to evolve more passively from the blue cloud onto the red sequence. With a typical run time of $\sim3$ minutes per value of $k$ for our galaxy sample, we show how $k$-means unsupervised clustering is an ideal tool for future analysis of large extragalactic datasets, being scalable, adaptable, and providing crucial insight into the fundamental properties of the local galaxy population.
\end{abstract}


\section{Introduction}
\label{sec:intro}

Understanding the diversity of galaxies in the Universe is a crucial part of understanding how galaxies form and evolve. Achieving this necessitates the use of a classification scheme to segregate galaxies in an astrophysically meaningful way. Galaxies have traditionally been morphologically classified using the Hubble sequence \citep{HUBBLE-1926,SANDAGE-2005}. Morphological classification schemes like the Hubble sequence describe a dichotomy of galaxies, distinguishing primarily between disk-dominated, clumpy, ``late'' type galaxies, and spheroid-dominated, smooth, ``early'' type galaxies. A significant success of morphological classifications is that they have enabled an understanding of the influence of the environment that a galaxy inhabits upon its evolution. The observed correlation of morphology with environment \citep{DRESSLER-1980,POSTMAN+1984} reveals how environmental processes like major mergers \citep{BARNES-1992,NAAB+2003}, minor mergers and tidal interactions \citep{TOOMRE+1972,PARK+2008}, and harassment \citep{MOORE+1996} change the structures of galaxies from late type to early type.

Dichotomies (or bimodalities) of galaxies have also been identified in star formation rate vs. mass \citep{SMETHURST+2015}, size vs. mass \citep{VANDERWEL+2014}, colour vs. size \citep{KELVIN+2014a}, S\'ersic index vs. colour, S\'ersic index vs. mass \citep{LANGE+2015}, star formation rate vs. S\'ersic index \citep{SCHIMINOVICH+2007}, and most commonly, colour vs. mass/magnitude \citep{BALDRY+2004,PENG+2010,TAYLOR+2015} feature spaces, among others. All of these bimodalities have been studied in detail individually, and many have been used to reveal the evolution of galaxies toward redder colours, higher masses, lower star formation rates, larger sizes, and higher S\'ersic indices over time. In particular, the colour vs. mass plane has revealed the evolution of galaxies from the blue sequence to the red sequence (e.g. \citealt{SALIM+2007}). That all of these bimodalities broadly align with one another and with the morphological bimodality suggests the existence of an overall, fundamental bimodality of galaxies in the Universe (e.g. \cite{DRIVER+2006}). In order to uncover the exact nature of this fundamental bimodality, any potential substructure thereof, and the evolutionary processes that drive it, all of these features (or a careful, representative selection thereof) must be considered together.

Meanwhile, approaches to the task of classifying samples of galaxies have evolved as sample sizes have increased over time. Historically, this task was reserved for expert astrophysicists who would use schemes like the Hubble sequence to morphologically classify the $10^{2-4}$ galaxies that would constitute a typical catalogue (e.g. \citealt{SANDAGE-1961, FUKUGITA+2007}). More recently, the Galaxy Zoo project (\citealt{LINTOTT+2008, LINTOTT+2011, WILLETT+2013}) crowd-sourced morphological classifications for $10^{5-6}$ galaxies from the Sloan Digital Sky Survey (\citealt{YORK+2000}). While both of these approaches have been useful, neither will scale to the data volume expected from the unprecedentedly large extragalactic surveys of the future (e.g. Euclid, from which photometry for $>10^{9}$ galaxies is expected; \citealt{LAUREIJS+2011}). Instead, fast and automated methods for galaxy classification will be required. There are two machine learning approaches that are suitable for this: supervised and unsupervised methods.

Much of the focus in the astrophysics literature has been on developing supervised classification methods, which assign unlabelled observations to a set of existing classes. In order to generate a model for this purpose, classification algorithms require training on pre-labelled observations. The model may also be validated using a second set of pre-labelled observations. Artificial neural networks have been proven to perform as well as expert observers in assigning Hubble type classifications \citep{LAHAV+1995, BALL+2004} based on sets of photometric features, and support vector machines have been shown to be able to distinguish between galaxy morphologies \citep{HUERTASCOMPANY+2008, HUERTASCOMPANY+2011} via the CAS morphological parameters (concentration, asymmetry, smoothness; \citealt{CONSELICE-2003}). Convolutional neural networks have been tested directly on pixel data \citep{DIELEMAN+2015, HUERTASCOMPANY+2015}.

The stronger the relationship between the classes and features, the more accurate the classification model. Ideally, classes are defined directly by the features, though this is not always the case. Schemes that depend on very large numbers of features, or whose dependence on features is poorly defined (e.g. the Hubble sequence), rely on the use of small sets of summary features (e.g. CAS) that correlate with their classes to act as a proxy. Workarounds like this often incur a loss of classification accuracy. Additionally, the accuracy of the classifications improves with the size and representativeness of the training sample, so large numbers of pre-labelled observations are required for the best results. A particular benefit of the use of supervised classification methods is the ease with which the accuracy of the classification model may be externally validated against pre-labelled observations.

Unsupervised clustering methods are less prevalent in the astrophysics literature. These methods are more exploratory in nature in that they search for groups of observations that are similar to one another, called clusters. Clustering algorithms do not use an existing classification scheme to segregate the observations and hence do not require training on pre-labelled observations. The clusters that are found may, however, be used as training labels for a supervised classification algorithm, thereby forming the basis of a new classification scheme. Previous examples of schemes generated in this way for the classification of galaxies involve large numbers of features and observations \citep{SANCHEZALMEIDA+2010, HOCKING+2017}, resulting in a large number of highly uniform classes. \cite{ELLIS+2005} apply a series of unsupervised methods to find a fundamental split of the galaxy population into two clusters which is dictated mostly by morphology. \cite{BARCHI+2016} use five morphological parameters to test clustering algorithms on their ability to find two distinct morphological classes. \cite{SIUDEK+2018} use a Fisher-Expectation-Maximisation algorithm (\citealt{BOUVEYRON+2012}; similar to the $k$-means method which is a simple Expectation-Maximisation algorithm) to define 12 different classes of galaxy via their absolute magnitudes in multiple bands and their redshifts.

Unsupervised clustering methods are attractive because they are descriptive. They model the structure of data directly without using any explicit prior notion of classes that might exist. Choice of features and algorithm will, however, implicitly influence the clusters that emerge. Features must offer sufficient discriminating information to the model while avoiding redundancies and overfitting. Algorithms vary in their definitions of a cluster and of similarity, and will find different clusters accordingly. Hence, it is necessary to select an appropriate combination of features and algorithm for a given clustering situation. Domain-specific knowledge (pertinent scientific theory) may also be used to inform these choices. Clusters are difficult to validate given the absence of truth labels to compare with. Instead, clusters are evaluated internally using some measure of cluster quality. Clusters may also be evaluated by comparison with other, existing classifications.

Limited by the need for visualisation, the data structure of the galaxy population (and the bimodality thereof in particular) has been well studied in feature spaces of up to three dimensions. An unsupervised clustering approach enables the study of the data structure of galaxies in feature spaces of higher dimensionalities, at which visualisation becomes more difficult. The specific approach we apply to do so is the $k$-means method, selected for its conceptual simplicity, speed, and popularity in the clustering literature \citep{JAIN+2010}. The method assumes that clusters are compact and spherical. We adopt an exploratory stance in our work, aiming to determine whether a higher dimensionality of feature space leads to previously unknown associations between features, whether these associations enable a deeper understanding of the fundamental bimodality of galaxies, and whether these associations engender a clustering structure in the feature data beyond the known bimodality of galaxies. We also aim to describe and explain this clustering structure in the context of current theories of galaxy formation and evolution.

We test the viability of the $k$-means method as a galaxy classification solution for the next generation of extragalactic surveys and as a tool to explore feature spaces of high dimensionalities using a redshift- and magnitude-limited sample of $7338$ galaxies from the GAMA survey. We represent our sample using a preliminary selection of five features. While our feature selection is preliminary and imperfect, we comment on how it influences the clusters that we find in our sample and how the selection might be improved for future studies. Cluster identities are discussed in terms of the input clustering features, by comparison with Hubble-like morphological classification, and in relation to the local environmental densities of the galaxies they contain.

The remainder of this paper is structured as follows. In section \ref{sec:km} (supplemented by appendix \ref{app:sim}) we describe our $k$-means implementation and our cluster evaluation method. In section \ref{sec:dat} we outline our sample, feature selection, and data preparation. In section \ref{sec:res} (supplemented by appendices \ref{app:bs} and \ref{app:ps}) we present and analyse clustering results, and in section \ref{sec:summ} we summarise, conclude, and consider directions for future work. Where required, we assume a ($H_{0}$, $\Omega_{m}$, $\Omega_{\Lambda}$) = ($70$ km s$^{-1}$ Mpc$^{-1}$, $0.3$, $0.7$) cosmology. We note that while the term ``cluster'' already has an established meaning in extragalactic astrophysics and cosmology, referring to groups of galaxies that are close in physical space in the Universe, we use it in this paper to refer to groups of galaxies that are close in feature space.

\section{\lowercase{$k$}-means and cluster evaluation}
\label{sec:km}

The $k$-means method is an unsupervised clustering approach that aims to partition a sample of $N$ observations, represented in a $D$-dimensional feature space, into $k$ compact, spherical clusters. Each of the clusters (a set of observations $C$) is characterised by its centroid ($\bf{\bar{c}}$); its arithmetic mean in each of the $D$ features. The standard $k$-means implementation (``\texttt{k-means}''; \citealt{MACQUEEN-1967, LLOYD-1982}) is a simple, fast algorithm comprising three steps:

\begin{enumerate}
\setcounter{enumi}{-1}
\item Initialise: $k$ initial centres are selected (e.g. uniformly at random) from the observations.
\item Assign: the observations are assigned to their nearest centre (by Euclidean distance); these assignments are clusters.
\item Update: the centroid of each assignment is calculated; these become the new, updated centres.
\end{enumerate}

Steps 1 and 2 are iterated until the algorithm converges; until there are no further differences between subsequent iterations. The convergence of \texttt{k-means} to a clustering solution is provably always finite \citep{SELIM+1984}, with a complexity $O(NDki)$ and generally requiring far fewer iterations ($i$) than there are observations \citep{DUDA+2000}. The final assignment of the observations may be taken as a classification scheme. The resultant partition of the sample is a Voronoi tessellation based on the final centroids. The final centroids are cluster archetypes: a $k$-point characterisation of the sample.

By iteratively recalculating the centroids, \texttt{k-means} inherently minimises the Sum of SQuare residuals Within ($SSQW$; equation \ref{eq:ssqw}; i.e. variance within) each of its clusters. The \texttt{k-means} definition of a cluster follows: clusters are data structures that are compact and separated (and therefore accurately characterised by their centroids), such that they have a lower $SSQW$. The total $SSQW$ of a set of \texttt{k-means} clusters, $\phi$ (equation \ref{eq:phi}), may be applied as an overall measure of their clustering quality. A consequence of the minimisation of $\phi$ is that \texttt{k-means} tends to produce clusters of similar sizes in feature space. This is called the ``uniform effect'' \citep{LIU+2010}, and it acts equally in all dimensions, leading also to spherical (rather than extended) clusters. It is common to normalise data to mitigate the influence of this effect on the results of \texttt{k-means}.

\begin{equation}
SSQW_{j} = \sum\limits_{\bf{c} \in C_{j}}||\bf{c}-\bf{\bar{c}}_{j}||^{2}.
\label{eq:ssqw}
\end{equation}

\begin{equation}
\phi = \sum\limits_{j = 1}^{k} SSQW_{j} = \sum\limits_{j = 1}^{k} \sum\limits_{\bf{c} \in C_{j}}||\bf{c}-\bf{\bar{c}}_{j}||^{2}.
\label{eq:phi}
\end{equation}

\texttt{k-means} is a local search heuristic: the behaviour of the centres as the algorithm iterates is dictated by those observations in their vicinities. Therefore, the outcome of \texttt{k-means} is dependent on the input initialisation. A common initialisation technique is to select $k$ observations from the sample uniformly at random. However, different such random initialisations may result in different, locally optimal clustering solutions. It is computationally impractical to search for the global optimum in all of the $k^{N}$ clustering permutations of a large sample. Hence, when presented with different solutions generated from different runs of \texttt{k-means} with the same $k$ on the same sample, it is standard practice to select as the optimal solution that with the lowest $\phi$.

To mitigate the local dependency of \texttt{k-means}, we apply the random initialisation technique of \cite{ARTHUR+2007} in \emph{all} of our runs of the algorithm. It spreads out the initial centres, making the subsequent results of \texttt{k-means} more competitive with globally optimal solutions. The first of the $k$ centres is selected from the sample with uniform probability. Subsequent centres are then selected with an increasing probability at larger distances from all preceding centres. This encourages optimisation to separated clusters. Whilst this initialisation is slower than a uniformly random initialisation, it generally yields a faster convergence over the iteration steps, resulting in a lower overall computation time for \texttt{k-means}.

\subsection{Stability}
\label{sec:stab}

A key consideration with applying \texttt{$k$-means} is the use of a suitable value of $k$; this is required as an input to the algorithm. \texttt{k-means} will always converge to a solution, even in the absence of clustering structure in the sample. Assuming the sample has a clustering structure, it is generally the case that $k_{true}$, the \emph{true} number of clusters in a given sample, is not known. It may even be that the true clusters in a sample have a hierarchical structure, such that there are several unknown values of $k_{true}$. Hence, it is common to trial clustering on a sample at several values of $k$ and identify good values for modelling the true clustering structure of the sample post-clustering. Comparing these results necessitates an additional, alternative measure of clustering quality; $\phi$ decreases systematically as $k$ increases because more clusters occupy the same sample.

We identify good values of $k$ based on the stability of their clustering solutions \citep{LISBOA+2013, LUXBURG-2010}. Specifically, we examine the stability of solutions in spite of random initialisations, which may result in different clustering solutions. Solutions at some values of $k$ may be more or less different to one another than solutions at other values of $k$. Those values of $k$ at which solutions are more similar to one another are more stable; \texttt{k-means} consistently converges to similar results, which implies a clustering structure in the sample at those values of $k$.

Stability may be understood by considering the behaviour of the \texttt{k-means} centres as the algorithm iterates. A key expectation is that if there is at least one centre in each of the $k_{true}$ clusters at initialisation, then the centres will remain within those true clusters as \texttt{k-means} proceeds \citep{BUBECK+2009}. For $k = k_{true}$, the centres will then settle to the centroids of the true clusters, in accordance with the algorithm's inherent minimisation of $\phi$. For $k > k_{true}$, this key expectation means that true clusters containing more than one centre at initialisation will be split. For $k < k_{true}$, where this key expectation does not hold, centres may move between true clusters and lead to mergers. The exact splits and mergers that occur are dependent on the locations of the centres at initialisation and will therefore change with different initialisations. It is important to note that when $k = k_{true}$, our choice of initialisation technique facilitates the ideal situation in which the $k_{true}$ clusters contain one centre each at initialisation. We demonstrate these concepts using a simple 2D simulation in appendix \ref{app:sim}.

To measure the difference between a pair of clustering solutions at the same $k$, we use Cram\'er's $V$ index of association \citep{CRAMER-1946}:

\begin{equation}
V = \sqrt{\frac{\chi^2}{N \cdot (k-1)}}.
\label{eq:v}
\end{equation}

Here, $\chi^2$ is the chi-squared value for two clustering solutions (categorical variables $A$ and $B$) on the same sample, each consisting of the same number ($k$) of unique labels. It is calculated (equation \ref{eq:chi}) using a $k \times k$ contingency table (a.k.a. cross tabulation), comparing the observed frequency of observations ($o$) in each cell ($a,b$) with its expected frequency ($e = N/k^{2}$; equal in every cell) given a null hypothesis of independence of the two solutions. We provide examples of contingency tables and the calculation of their corresponding $\chi^{2}$ and $V$ values in appendix \ref{app:sim}.

\begin{equation}
\chi^{2}_{A,B} = \sum\limits_{a,b} \frac{(o_{a,b} - e_{a,b})^2}{e_{a,b}}.
\label{eq:chi}
\end{equation}

Cram\'er's $V$ index is normalised, reporting $\chi^{2}$ as a square-root-scaled fraction of its maximum possible value given $N$ (the number of observations in the sample) and $k$. It ranges from $0$ for no agreement (i.e. the solutions are independent; agreement is consistent with uniform random chance) to $1$ for perfect agreement. In practice, \texttt{k-means} cannot produce solutions that disagree to the extent that $V = 0$. We assess the stability of an individual clustering solution by calculating its median $V$ with respect to other solutions at the same $k$. We assess the stability of a set of clustering solutions at the same $k$ by examining their distribution in median $V$ (see figures \ref{fig:stab}, \ref{fig:simstab}, and \ref{fig:bsstab}). Stability also enables us to determine whether there is no clustering structure to our sample (i.e. no $k_{true}$), as no particular value of $k$ will stand out from the others as being particularly stable.

\section{Data}
\label{sec:dat}

We use data from phase II of the Galaxy And Mass Assembly (GAMA) survey \citep{DRIVER+2009, LISKE+2015}. The main aim of the survey is to study cosmic structure on scales of $1$kpc to $1$Mpc in the context of cold dark matter models of the Universe. The survey is structured around its spectroscopic campaign, conducted at the Anglo-Australian Telescope using the AAOmega spectrograph, and based on an input catalogue defined by \cite{BALDRY+2010}. The spectroscopy provided reliable heliocentric redshifts for $238000$ objects to a limiting r-band Petrosian magnitude of $19.8$ and across five regions covering a total area of $286$ deg$^{2}$. It has been supplemented with reprocessed imaging in 21 bands from a variety of other surveys (e.g. the Sloan Digital Sky Survey; \citealt{YORK+2000}) that overlap with the GAMA spectroscopic campaign footprint (the Panchromatic Data Release; \citealt{DRIVER+2016}). Data derived from these spectra and images are listed in tables hosted at \url{http://www.gama-survey.org/}.

We derive our sample from the well-characterised sample of \cite{MOFFETT+2016} (listed in the GAMA survey table VisualMorphologyv03) with a view to facilitating the interpretation of clustering results, particularly by comparison with results from other, previous GAMA survey studies. It is a flow-corrected-redshift- ($0.002 < z < 0.06$) and magnitude- ($r_{PETRO} < 19.8$) limited sample of $7556$ local objects that have been morphologically classified using the method of \cite{KELVIN+2014a}. Note that while we aim to compare clustering results with these visual, Hubble-like morphologies (see section \ref{sec:res}), we do not aim to reproduce them. 

The \cite{KELVIN+2014a} method assigns classifications by the consensus of three expert observers, whose visual inspection of optical three-colour images of the galaxies is guided by a decision tree. The tree discriminates galaxies firstly as being either spheroid- or disk-dominated, and secondly as consisting either of a single component or of multiple components. The tree goes on to discern multiple component galaxies with bars from those without, but we ignore this distinction in our study due to the relatively low numbers of barred galaxies in our sample ($\sim 4$ per cent of all of the galaxies have bars). We intend to examine bars in the context of our clusters in a future study, however, due to the significant role they play in the evolution of any barred galaxy, particularly in terms of quenching star formation (e.g. \citealt{SHETH+2005}). 

The two levels of the tree that we do consider lead to four morphological types: E, S0-a, Sab-Scd, and Sd-Irr. A fifth type, ``Little Blue Spheroid'' (LBS), is identified separately at the top level of the tree. As star-forming, spheroid-dominated, blue dwarf galaxies, they have been likened to the blue early type galaxies of \cite{SCHAWINSKI+2009} in that they defy typical galaxy trends with colour or morphology. Contaminants are also identified at the top level of the tree. There are $25$ in the sample, which are mostly satellite galaxies, partial galaxy structures, or star-galaxy blends. We remove them from the sample, leaving $7531$ galaxies.

\begin{table*}
\caption{The features we use to characterise galaxies, and the survey tables from which we retrieve them.}
\label{tab:feats}
\centering
\begin{tabular}{l l l r l l}
\hline
Feature        					& Unit						& Table 				& Version 	& Column							& Reference 				\\
\hline
Stellar mass 					& $\mathrm{M_{\odot}}$		& MagPhys 				& $6$ 		& mass{\_}stellar{\_}best{\_}fit	& \citealt{DRIVER+2016}	\\
u-r colour						& mags						& StellarMassesLambdar	& $20$		& uminusr							& \citealt{TAYLOR+2011}	\\
S\'ersic index					& - 						& SersicCatSDSS 		& $9$		& GALINDEX{\_}r						& \citealt{KELVIN+2012}	\\
Half-light radius				& kpc						& SersicCatSDSS 		& $9$		& GALRE{\_}r						& \citealt{KELVIN+2012}	\\
Specific star formation rate	& $\mathrm{yr^{-1}}$		& MagPhys				& $6$		& sSFR{\_}0{\_}1Gyr{\_}best{\_}fit	& \citealt{DRIVER+2016}	\\
\hline
\end{tabular}
\end{table*}

We then retrieve feature data for the galaxies in the sample. There are hundreds of features available in the GAMA database with which to characterise the galaxies in the sample. One may be tempted to find clusters in the sample using all of them at once, in order to ``provide the algorithm with as much information as possible''. However, as the dimensionality of the feature space containing the sample increases, the observations become more sparse, and \texttt{k-means} (or any clustering algorithm) overfits its clusters to the observations. Representing the sample using a smaller subset of features instead results in clusters that are more readily generalisable to the overall galaxy population. 

Feature selection may involve both domain-specific knowledge and statistical considerations. We select features that capture intrinsic properties of galaxies, relating to their formation and evolution, and aim for a selection that expresses most known aspects of these processes. The redundancy of features - the extent to which they provide the same information as other features - may be assessed statistically, such as by calculation of their Spearman rank-order correlation coefficients. A higher correlation between features implies greater redundancy. Redundant features exert a higher influence over the clusters that \texttt{k-means} finds in that they lead to a projection of the feature space in which the sample is highly extended in one direction over others (e.g. figure \ref{fig:pca}). \texttt{k-means} will therefore tend to split the data along the extended direction of the data due to the uniform effect. While it is common to discard such features to avoid this bias, retaining them could instead serve to strengthen a desirable pattern in the data.

Table \ref{tab:feats} lists the feature data we retrieve and the main GAMA survey tables we access to do so. Stellar masses ($M_{*}$) and gigayear-timescale specific star formation rates ($SSFR$) are taken from MagPhysv06, generated by \cite{DRIVER+2016} from a run of \texttt{MAGPHYS} \citep{DACUNHA+2008} on all 21 bands of foreground extinction-corrected photometry listed in LambdarCatv01 \citep{WRIGHT+2016}. \texttt{MAGPHYS} calculates spectral energy distributions (SEDs) from input redshifts, fluxes, and flux errors using star and dust emission template spectra, and corrects for light-attenuation by dust within galaxies. Restframe $u-r$ colours come from StellarMassesLambdarv20, derived from the same photometry \citep{TAYLOR+2011}. We select $u-r$ colour for its ability to express the galaxy bimodality in the colour vs. mass plane. Unlike $M_{*}$ and $SSFR$, it does not include corrections for dust attenuation, meaning our clustering results will be influenced in some capacity by the presence of dust in some of the galaxies in our sample. We take $r$-band S\'ersic indices ($n$) and half-light radii ($R_{1/2}$) from SersicCatSDSSv09, whose derivation is described in \cite{KELVIN+2012} and is based on reprocessed Sloan Digital Sky Survey imaging \citep{YORK+2000, HILL+2011}. The accuracy of S\'ersic indices is expected to be consistent throughout our sample due to the low redshifts of the galaxies therein \citep{VIKA+2013}. 

Matching for data in all features leaves $7516$ galaxies in the sample, with 15 lost due to incompleteness. We note that this feature selection is preliminary. Our use of features derived primarily from broad-band photometry facilitates a comparison of clustering results with a wide range of other surveys. We comment on the consequences of our feature selection for clustering later in this section and in section \ref{sec:res}, and the potential for optimising feature selection in section \ref{sec:summ}.

The half-light radii listed in SersicCatSDSSv09 are presented in units of arcseconds, which are a function of the distances to the galaxies as well as their intrinsic sizes. We use flow-corrected redshifts (DistanceFramesv12; \citealt{BALDRY+2012}) to convert them to intrinsic kiloparsec radii.

We apply a series of transforms to standardise the data, with the intention of avoiding unintuitive partitions due to the uniform effect, and of granting equal weight to all of the features. The distributions of all of the features except $u-r$ colour are strongly skewed. The centroids that \texttt{k-means} iteratively recalculates are sensitive to the uneven tails of skewed distributions. We therefore ensure that all features are represented in logarithmic units (see table \ref{tab:lim}). Additionally, these features are all commonly represented in logarithmic units in the astrophysics literature.

\begin{table}
\caption{The limits for truncation that have been imposed on each of the features. The truncated histograms are viewable in figure \ref{fig:samp}. Limits marked with an asterisk do not actually exclude any galaxies.}
\label{tab:lim}
\centering
\begin{tabular}{l l r r}
\hline
Feature     & Units 					& Lower		 & Upper		\\
\hline
$M_{*}$		& $\log_{10}$(M$_{\odot}$)	& $6$ 		& $^{\ast}12$	\\
$u-r$		& mags						& $0.3$		& $2.7$			\\
$n$			& $\log_{10}$(n)			& $-0.6$	& $1.2$			\\
$R_{1/2}$	& $\log_{10}$(kpc) 			& $-1.0$	& $^{\ast}1.5$	\\
$SSFR$		& $\log_{10}$(yr$^{-1}$)	& $-14$		& $-8$			\\
\hline
\end{tabular}
\end{table}

Outliers in the sample are more readily apparent when examining the distributions of each of the features in logarithmic units. In order to mitigate the influence of outliers on the calculation of centroids by \texttt{k-means}, we truncate the sample in each of the features, removing galaxies that lie outside given limits. The limits we impose are listed in table \ref{tab:lim}. While some of these limits have been set simply by inspection of histograms of the sample in each of the features, others have also involved astrophysical considerations. For example, the limits in S\'ersic index have been set to eliminate subcomponent fits, or fits affected by light from sources near the galaxy that was fitted. Removing $198$ outliers from the sample leaves $7338$ galaxies. 

\begin{figure*}
\centering
\includegraphics[width=0.9\textwidth]{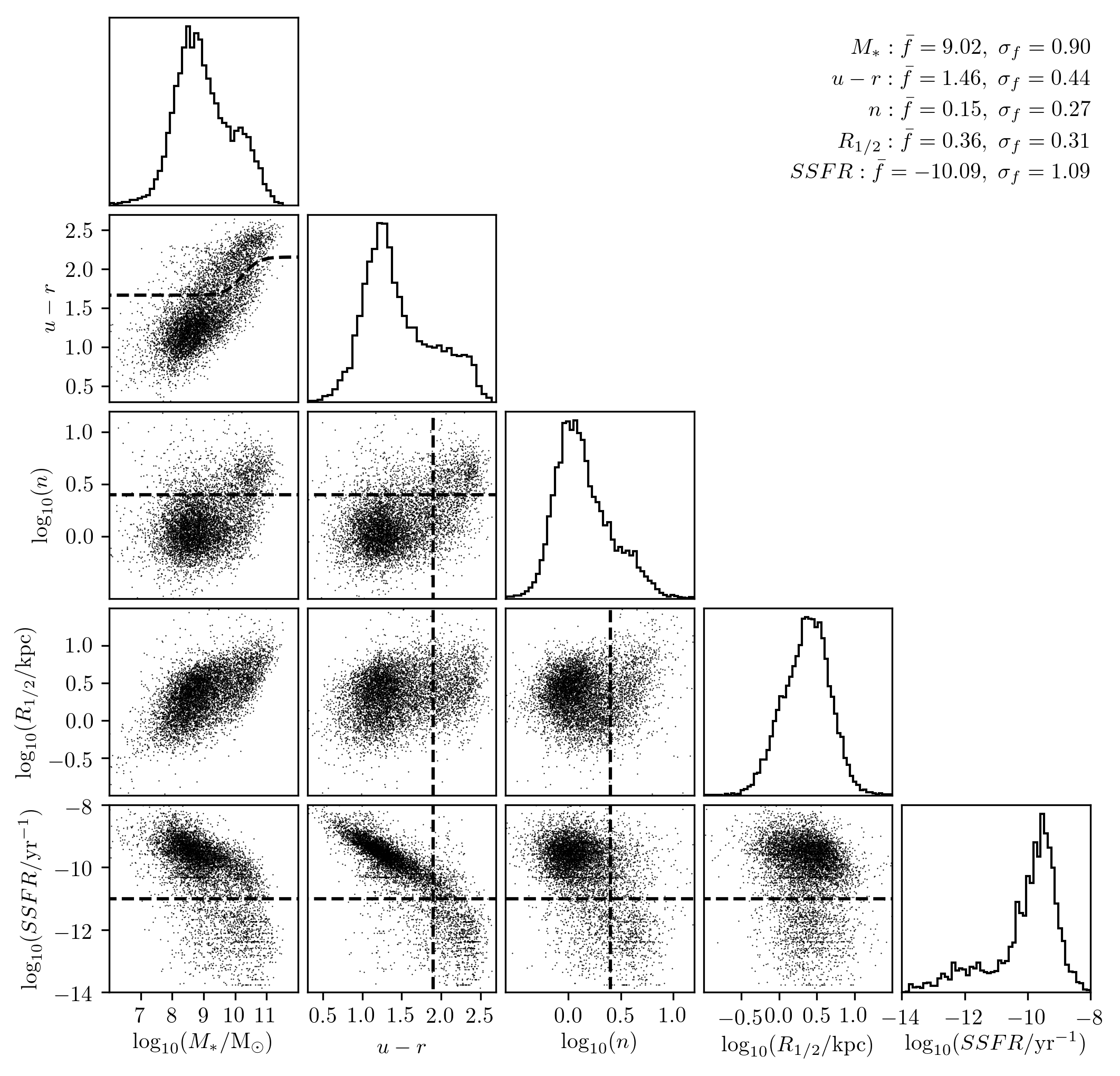}
\caption{A profile of our sample. It is represented using histograms and scatter plot projections. The dashed lines mark ``classical'' distinctions between the two main populations of galaxy (see text). We also list the mean ($\bar{f}$ and standard deviation ($\bar{\sigma}$) of the sample in each feature in the top-right of the figure, in the units shown on the axes.}
\label{fig:samp}
\end{figure*}

This now constitutes our final sample, which we profile in figure \ref{fig:samp}. We use histograms and scatter plot projections to show the distribution of our sample in 1D and 2D feature spaces. These distributions reveal a dominance of low mass, blue, star-forming galaxies with low S\'ersic indices in our sample. Most features exhibit significant secondary components to their distributions. The bimodality of galaxies is visible in several of the scatter plot projection panels. 

We include dashed lines in figure \ref{fig:samp} that mark ``classical'' distinctions from the literature that have been made between the two main populations of galaxies. The line in the colour-mass panel is based on equation 11 of \cite{BALDRY+2004}. We use a solar r-band absolute magnitude of $4.71$ and equation 12 from the same paper to adapt it from applying to magnitudes to applying to masses. While the original line was calculated using SDSS model magnitudes, the \texttt{LAMBDAR} apertures were set using \texttt{Source Extractor} (\citealt{BERTIN+1996}; see also section 6.1 of \citealt{WRIGHT+2016}). Model magnitudes report redder colours than magnitudes derived using tophat apertures: we calculate an approximate mean offset of $-0.15$ over the range of colours in our sample and adjust the line accordingly. The lines in the scatter panels involving S\'ersic indices and $u-r$ come from \cite{LANGE+2015}. We apply a similar colour offset of $+0.4$ as the $u-r$ colours in \cite{LANGE+2015} are corrected for dust attenuation. The line for $SSFR$ is taken from \cite{POZZETTI+2010}; specifically, we take their distinction between passive and non-passive galaxies.

\begin{figure}
\centering
\includegraphics[width=0.45\textwidth]{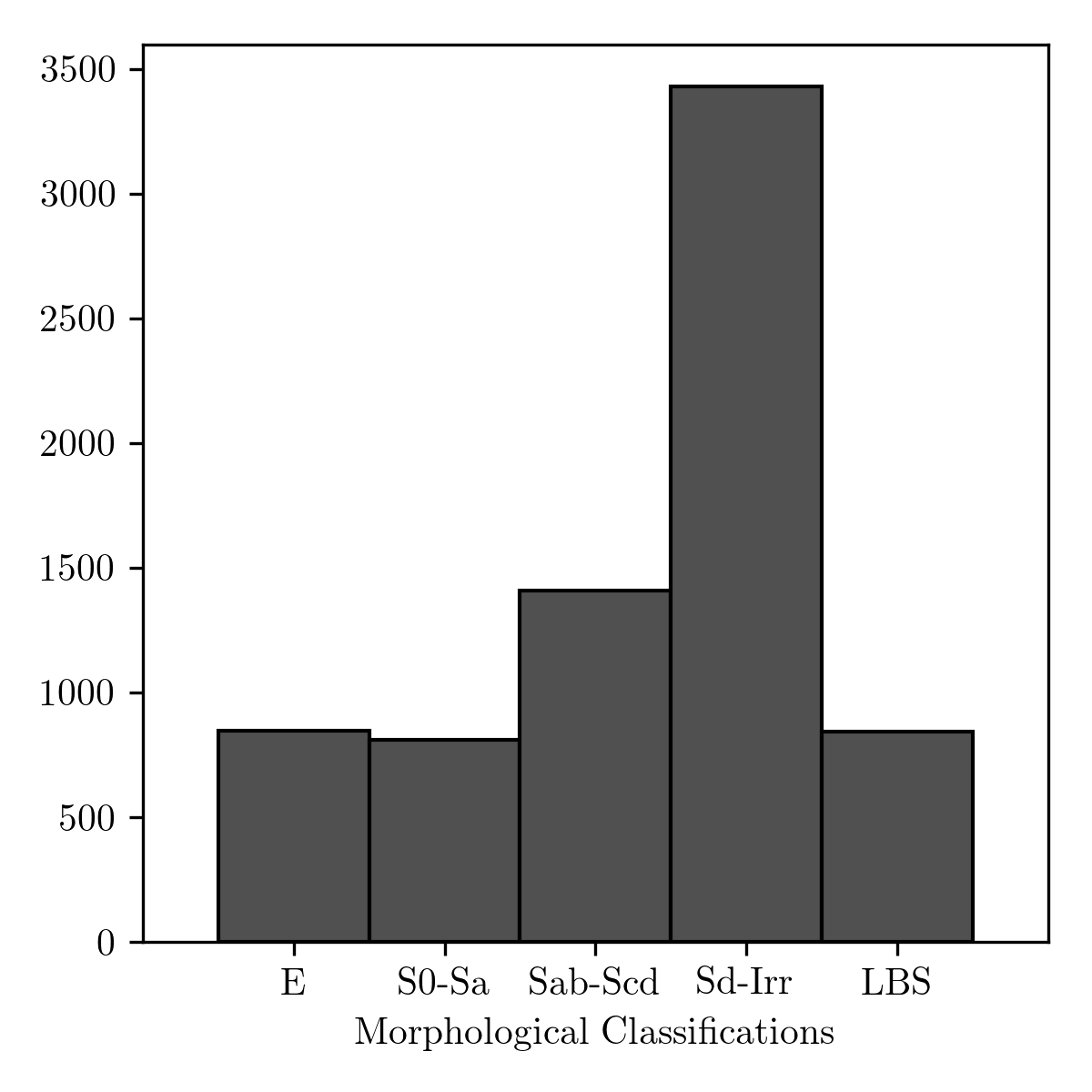}
\caption{Histogram showing the distribution of \protect\cite{KELVIN+2014a} and \protect\cite{MOFFETT+2016} morphologies in our sample. Our sample is dominated by late-type morphologies. LBS stands for ``Little Blue Spheroid''.}
\label{fig:h}
\end{figure}

\begin{figure}
\centering
\includegraphics[width=0.45\textwidth]{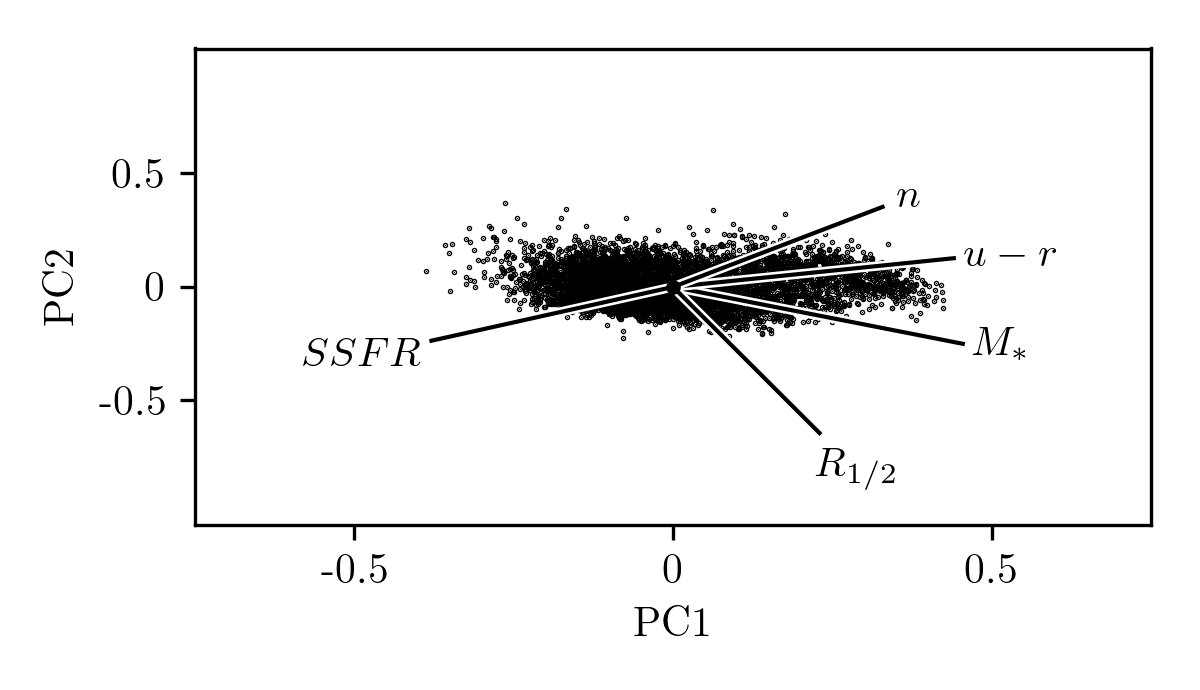}
\caption{Our features as functions of the first two principal components of our sample. The axes are scaled to show the relative variance that each principal component encompasses. We also represent our sample using a scatter plot projection. The size of the sample in this 2D principal component space is normalised to fit within the area shown. $M_{*}$, $u-r$, and $SSFR$ are most strongly associated with PC1, while $R_{1/2}$ is most strongly associated with PC2. $n$ is evenly balanced between the two, but is most strongly associated with PC3, which we do not show.}
\label{fig:pca}
\end{figure}

Figure \ref{fig:h} shows the distribution of \cite{KELVIN+2014a} morphologies in our sample. The histogram reveals a dominance of late-type morphologies in our sample. The morphological types are ordered from highest (E) to lowest (LBS) mean mass.

Some features span larger numerical ranges in logarithmic units than the others. For example, $M_{*}$ spans $6$ orders of magnitude (base 10) while $n$ spans $1.8$. Given that \texttt{k-means} minimises $\phi$ in all dimensions, it will tend to split our sample along any direction in which it is extended. To mitigate any bias of \texttt{k-means} for or against any of our features on this basis, we code the data in each of the features using $Z$-scores. They are more strongly influenced by the centres of the feature distributions than their extremities (which normalisation techniques alternatively are). Hence, they weighted toward the majority of galaxies near the feature distribution means, rather than the minority of outliers. Here (equation \ref{eq:z}), $f$ is the value of an observation in a given feature, $\bar{f}$ is the mean value of that feature, $\sigma_{f}$ is its standard deviation, and $Z_{f}$ is the $Z$-score of $f$:

\begin{equation}
\label{eq:z}
Z_{f} = \frac{f - \bar{f}}{\sigma_{f}}.
\end{equation}

Having been standardised, we assume that \texttt{k-means} will now be able to recover clustering structure in our sample that reflects the astrophysics involved in the formation and evolution of the galaxies therein. We assess our feature selection pre-clustering. In table \ref{tab:sp} we show the Spearman rank-order correlation coefficients for pairs of the five features we use to represent our sample. We note that $u-r$ colour is involved in the two strongest correlations of features (with $M_{*}$ and $SSFR$) for our sample, suggesting redundancy. We opt to retain it, however, to strengthen the bimodal structure of the data and because it includes information about the dust content of the galaxies in our sample, which $SSFR$ does not. We expect that strengthening the bimodality will encourage \texttt{k-means} to search for more clusters \emph{within} the two peaks of the bimodality at higher values of $k$. Other correlations that exist among our selection of features are weaker and do not suggest any further significant redundancies.

\begin{table}
\caption{Spearman rank-order correlation coefficients for the features we use to represent our sample.}
\label{tab:sp}
\centering
\begin{tabular}{l | r r r r r}
\hline
Feature		& $M_{*}$	& $u-r$ 	& $n$		& $R_{1/2}$	& $SSFR$	\\
\hline
$M_{*}$ 	& $1.00$	&			& 			& 			& 			\\
$u-r$		& $0.72$	& $1.00$	&			&			&			\\
$n$		 	& $0.35$	& $0.40$	& $1.00$	& 			& 			\\
$R_{1/2}$	& $0.55$	& $0.25$	& $-0.04$	& $1.00$	& 			\\
$SSFR$		& $-0.61$ 	& $-0.83$	& $-0.38$	& $-0.19$	& $1.00$	\\
\hline
\end{tabular}
\end{table}

We also conduct a principal component analysis of our sample, to gain insight into its covariance structure and to anticipate clustering results. We list the results of this analysis in table \ref{tab:pca}. The results reveal that the structure of our sample is dominated by the first principal component (PC1), which encompasses $59$ per cent of our sample's variance. This tells us that the sample has an elongated shape in the five-dimensional feature space. PC1 is defined mostly by those features that reflect aspects of the stellar populations within galaxies (i.e. $M_{*}$, $u-r$, and $SSFR$), so we expect that these features will most strongly dictate the clusters that \texttt{k-means} finds. $R_{1/2}$ and $n$ are most strongly associated with PC2 (encompassing $21$ per cent of the variance in the sample) and PC3 ($12$ per cent) respectively. We expect them to play a role in dictating clusters at higher values of $k$, at which the use of additional centroids enables the algorithm to explore subtler, more local structure within our sample. These relationships are clearly apparent in figure \ref{fig:pca}, which shows our features and sample as functions of the first two principal components.

\begin{table}
\caption{Results of a principal component analysis of our sample.}
\label{tab:pca}
\centering
\begin{tabular}{l r r r r r}
\hline
Feature				& PC1		& PC2 		& PC3		& PC4		& PC5		\\
\hline
$M_{*}$ 			& $0.51$	& $-0.28$	& $-0.02$	& $-0.63$	& $-0.51$	\\
$u-r$				& $0.53$	& $0.15$	& $-0.32$	& $-0.22$	& $0.74$	\\
$n$		 			& $0.37$	& $0.40$	& $0.84$	& $0.07$	& $0.04$	\\
$R_{1/2}$			& $0.29$	& $-0.80$	& $0.21$	& $0.44$	& $0.19$	\\
$SSFR$				& $-0.49$ 	& $-0.30$	& $0.39$	& $0.18$	& $0.40$	\\
\hline
Relative Variance 	& $0.59$	& $0.21$	& $0.12$	& $0.05$	& $0.03$	\\
\hline
\end{tabular}
\end{table}

Finally, we also retrieve environmental data for the galaxies in our sample in order to probe the role of environment in dictating the clusters we find via its influence on our features. We adopt the surface density $\Sigma_{5}$, defined using the projected comoving distance from a galaxy to its fifth nearest neighbour, as a measure of local environmental density (via EnvironmentMeasuresv05; \citealt{BROUGH+2013}). This feature is only available for $4195$ of the $7338$ galaxies in our sample if we filter for a SurfaceDensityFlag of $0$, which ensures that the fifth nearest neighbour of a given galaxy lies within the GAMA survey footprint. Nearly all of the other $3143$ galaxies in our sample have a SurfaceDensityFlag of $2$, indicating no neighbours within a velocity cylinder of $\pm 1000$ km s$^{-1}$, meaning they occupy particularly low density environments.

We note that these $3143$ galaxies are not evenly distributed in feature space. Most are blue, low mass, and have low S\'ersic indices, large radii, and high specific star formation rates, consistent with the \cite{GOMEZ+2003}, \cite{BAMFORD+2009}, and \cite{PENG+2010} findings that such galaxies tend to occupy lower density environments. We comment on the consequences of this incompleteness for our examination of local environmental densities within clusters where relevant in section \ref{sec:res}. Naturally, our confidence in the conclusions we come to based on this data would be greater were this data available for the entirety of our sample.

\begin{figure}
\centering
\includegraphics[width=0.45\textwidth]{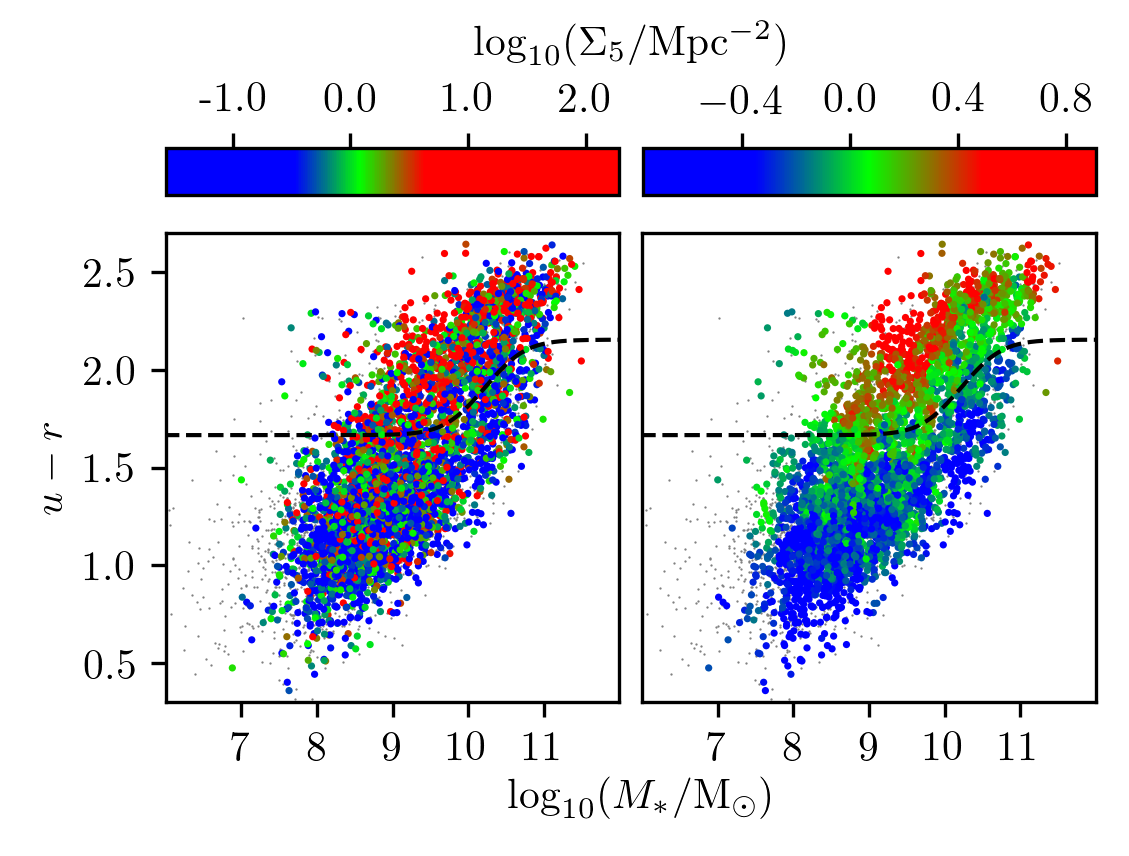}
\caption{Our sample, projected onto the $u-r$ vs. $M_{*}$ plane, with points coloured by their measured (left panel) and smoothed (right panel) local environmental densities ($\Sigma_{5}$). The small grey points represent those galaxies for which $\Sigma_{5}$ is not available; their bias toward lower masses and bluer colours is apparent. The dashed black line marks the \protect\cite{BALDRY+2004} distinction between the blue and red sequences of galaxies.}
\label{fig:e}
\end{figure}

Figure \ref{fig:e} shows our sample projected onto the $u-r$ vs. $M_{*}$ plane. In the left panel, points are coloured by their measured $\Sigma_{5}$, taken directly from EnvironmentMeasuresv05. In the right panel, the local environmental densities have been smoothed using a nearest-neighbour-averaging algorithm. We apply this smoothing to capture the average trend of environment with colour and mass for later analyses. We note that this smoothing inhibits the range of $\Sigma_{5}$ for our sample in the right panel compared with the left. The colour bar levels have been set in order to distinguish galaxies in intermediate density environments from those in low and high density environments. Both panels show that our sample is dominated by galaxies in low density environments. Interestingly, we note that the \cite{BALDRY+2004} dashed black line, intended as a separator of the blue and red sequences of galaxies in this feature plane, traces intermediate densities.

\section{Results}
\label{sec:res}

We combine stability and compactness (see section \ref{sec:km}) to evaluate \texttt{k-means} clusters in our sample, adapting the approach of \cite{LISBOA+2013}. We do not assume a value of $k_{true}$ (other than $k = 2$, corresponding to the bimodality of galaxies), so we trial \texttt{k-means} clustering at $k = 2$ through $k = 15$, initialising $200$ times at each $k$ using the \cite{ARTHUR+2007} technique. We first identify stable values of $k$, at which there appears to be a clustering structure in our sample, using $V$ (a measure of the strength of association between two clustering solutions; equation \ref{eq:v}). We then select the optimal solution at each of our stable values of $k$ by considering compactnesses.

\begin{figure}
\centering
\includegraphics[width=0.45\textwidth]{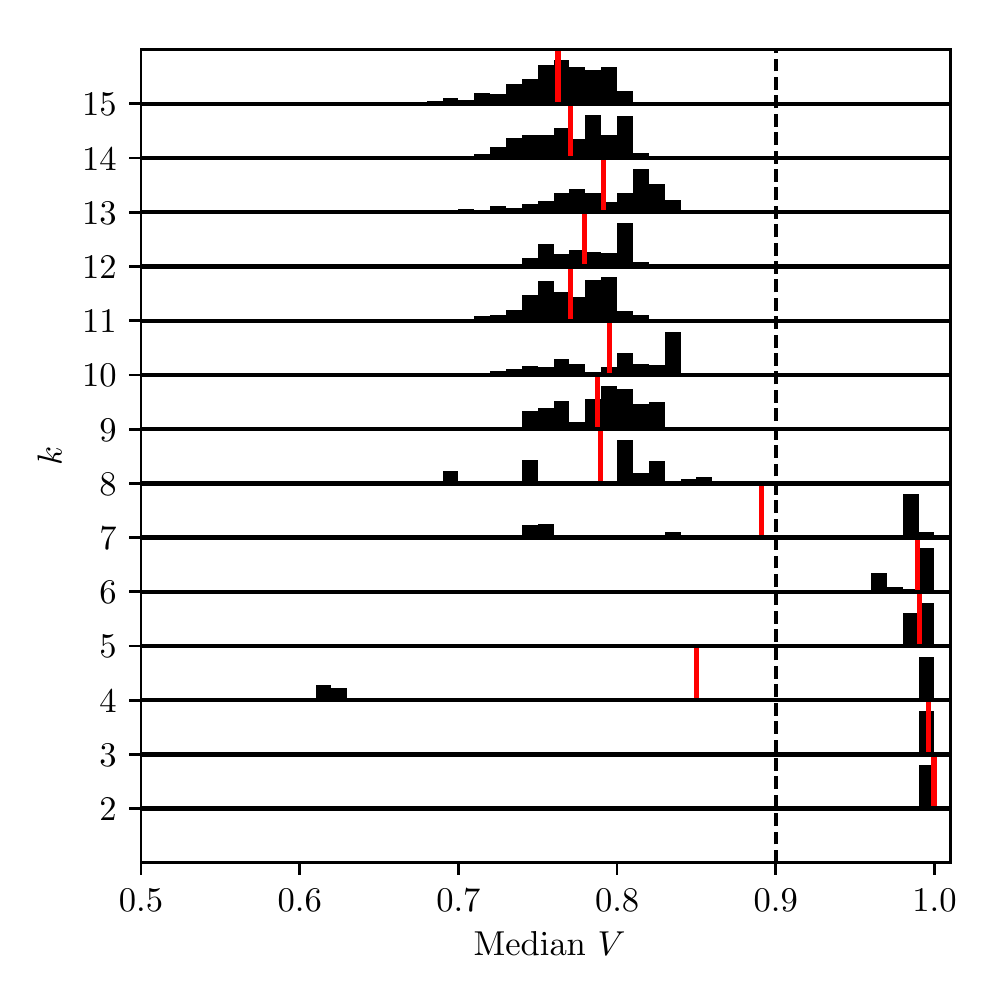}
\caption{Stability map of \texttt{k-means} clustering for our sample at $k = 2$ through $k = 15$. We calculate the median $V$ of each solution with respect to all other solutions at the same $k$. The distributions of all $200$ medians at each $k$ are represented using histograms plotted along each of the horizontal black baselines. The heights of the histograms are normalised. Additionally, we show the means of these distributions as vertical red lines. Solutions at $k = 2, 3, 5,$ and $6$ are particularly stable.}
\label{fig:stab}
\end{figure}

In figure \ref{fig:stab} we map the stabilities of solutions at different values of $k$. We calculate the median $V$ of each individual solution with respect to all other solutions at the same $k$. The distributions of all $200$ medians at each $k$ are represented using histograms plotted along each of the horizontal black baselines. The heights of the histograms are normalised. Additionally, we show the means of these distributions as vertical red lines. We note a gap across all distributions, appearing to separate two distinct regimes of solutions. We demarcate these regimes using the vertical dashed black line at median $V = 0.9$, but emphasise that this ``threshold'' is a product of our sample and may differ for different samples. The key element for distinguishing between stable and unstable values of $k$ is the gap.

Values of $k$ at which the distributions are concentrated toward higher median $V$ (i.e. at which more solutions are more consistent) are more stable. The solutions at $k = 2, 3, 5,$ and $6$ stand out as being particularly stable. All of the solutions at each of these values of $k$ have median $V > 0.9$, except for a single solution at $k = 3$. The spread of solutions at $k = 6$ corresponds to a maximum difference of $\sim 100$ galaxies ($\sim 1.5$ per cent of our sample) between solutions.

The solutions at $k = 4$ occupy a highly-peaked bimodal distribution, with a slight majority ($123$) at median $V > 0.9$. The solutions at $k = 7$ are similarly distributed with $113$ at median $V > 0.9$, but with a larger spread in its secondary peak. While both distributions exhibit stable components, a significant number of solutions in each are unstable. We focus presently on those values of $k$ that are most uniformly stable, and therefore exclude any solutions from $k = 4$ and $k = 7$ from our analyses of clustering results.

The distributions of solutions at higher values of $k$ are centred at lower median $V$ and have larger spreads, meaning they are unstable. The additional centroids used by \texttt{k-means} at these higher values of $k$ are more strongly influenced by the local structure within our sample at initialisation, such that the algorithm is more likely to converge to locally (rather than globally) optimal solutions. The ability of our initialisation technique to mitigate the local dependency of \texttt{k-means} becomes weaker as $k$ increases. The spreads of the solutions for these unstable values of $k$ correspond to differences of thousands of galaxies between solutions. We note that the general trend of decreasing stability at higher $k$ continues beyond the solutions at $k = 15$.

From each of the four values of $k$ that we have identified as being most stable, we select as our final, optimal solutions for analysis those with the lowest $\phi$ (a measure of the compactness of the clusters in a solution; equation \ref{eq:phi}). We refer to these four ``best'' solutions as simply $k = 2$, $k = 3$, $k = 5$, and $k = 6$. We note that these solutions at these values of $k$ retain their stability following application of the bootstrap method to our sample (appendix \ref{app:bs}). We note also that the stability map that results from clustering for which $u-r$ colour is omitted as an input feature has the same general structure as figure 5, but the distributions of solutions are systematically offset to slightly lower median $V$ at each value of $k$.

\begin{figure}
\centering
\includegraphics[width=0.45\textwidth]{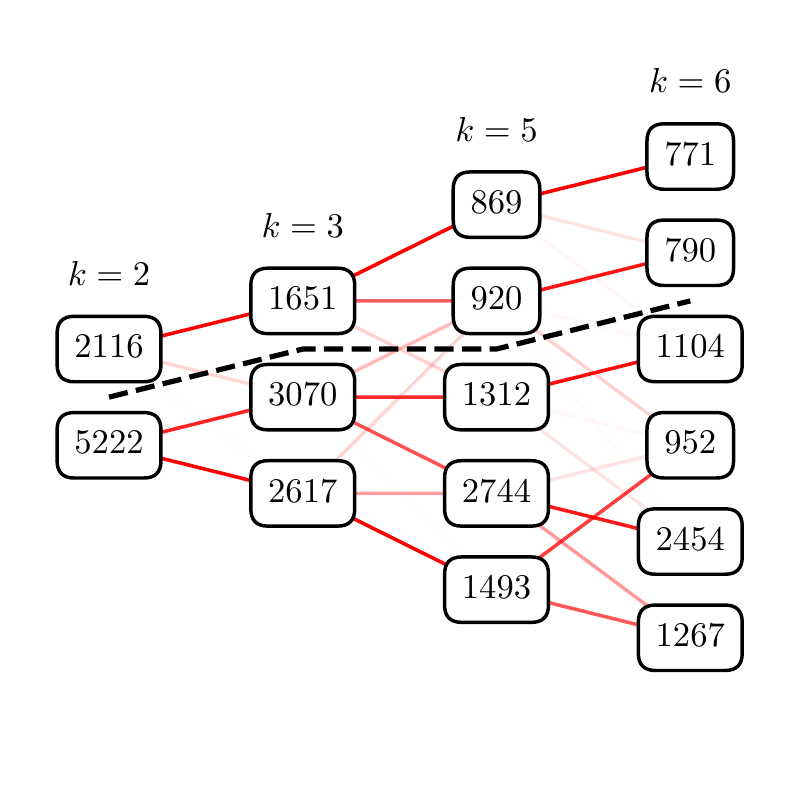}
\caption{Hierarchy tree showing the interrelation of $k = 2$, $k = 3$, $k = 5$, and $k = 6$. The text bubbles, representing the clusters, state the number of galaxies they contain and are ordered by the clusters' mean $u-r$ colours from reddest at the top to bluest at the bottom. The opacity of the red lines expresses how closely related connected clusters at different $k$ are. The dashed black line separates the basic structure of two ``superclusters'' that we find at all values of $k$.}
\label{fig:hi}
\end{figure}

A hierarchy tree, mapping the interrelation of our best solutions, is shown in figure \ref{fig:hi}. The clusters in each solution are represented by text bubbles which state the number of galaxies that they contain. The red lines express how closely related clusters at different $k$ are. The opacities of the lines scale linearly with the fraction of galaxies in clusters at $k + 1$ that are also found in clusters at $k$. Clusters are ordered vertically at each $k$ by the mean $u-r$ colour of the galaxies they contain, with the reddest clusters at the top and the bluest at the bottom. 

It should be noted that solutions at different values of $k$ are calculated independently of one another, so hierarchy is not imposed or assumed at any point in the clustering. Despite this, we find that our best solutions exhibit a broadly hierarchical structure. Considering them in sequence, clusters at higher values of $k$ generally emerge as splits of clusters at lower values of $k$. There is some mixing present, meaning some clusters at higher values of $k$ contain galaxies from multiple clusters at lower values at $k$. This is especially noticeable between $k = 3$ and $k = 5$, though it may be exaggerated due to the omission of a solution at $k = 4$ from the plot. The highly peaked bimodal distribution of solutions in median $V$ at $k = 4$ (figure \ref{fig:stab}) arises as $k$-means settles into one of the two splits that must occur between $k = 3$ and $k = 5$. $k = 5$ is stable and includes both of these splits, so no information is lost by the exclusion of a solution at $k = 4$ from our analyses.

\begin{table*}
\renewcommand{\arraystretch}{1.5}
\caption{A summary of all of the clusters in solutions $k = 2, 3, 5,$ and $6$. See the main text for an explanation of cluster names. The uncertainties on the centroids are estimated by application of the bootstrap method to our sample; we outline this estimation in detail in appendix \ref{app:bs}.}
\label{tab:c}
\centering
\begin{tabular}{l r | r r r r r | r r}
\hline
Cluster	& $N_{C}$ 	& $\log_{10}(M_{*}/$M$_{\odot})$	& $u-r$ 						& $\log_{10}(n)$					& $\log_{10}(R_{1/2}/$kpc$)$	& $\log_{10}(SSFR/$yr$^{-1})$	& $\log_{10}(\Sigma_{5}/$Mpc$^{-2})$			& Loss \%	\\
\hline
Ra$_{2}$	& $2116$	& $10.022_{-0.006}^{+0.007}$	& $2.018_{-0.004}^{+0.004}$		& $0.373_{-0.003}^{+0.003}$		& $0.506_{-0.001}^{+0.002}$	& $-11.35_{-0.01}^{+0.01}$	& $0.16$	& $30$	\\
Ba$_{2}$	& $5222$	& $8.611_{-0.004}^{+0.004}$		& $1.240_{-0.002}^{+0.002}$		& $0.0565_{-0.0002}^{+0.0004}$	& $0.300_{-0.001}^{+0.001}$	& $-9.579_{-0.003}^{+0.004}$	& $-0.21$	& $48$	\\
\hline
Ra$_{3}$	& $1651$	& $10.072_{-0.011}^{+0.007}$	& $2.101_{-0.011}^{+0.005}$		& $0.445_{-0.008}^{+0.005}$		& $0.472_{-0.002}^{+0.002}$	& $-11.64_{-0.02}^{+0.04}$		& $0.23$	& $30$	\\
Bb$_{3}$	& $3070$	& $9.14_{-0.04}^{+0.02}$		& $1.391_{-0.018}^{+0.009}$		& $0.035_{-0.007}^{+0.005}$		& $0.528_{-0.005}^{+0.006}$	& $-9.80_{-0.02}^{+0.03}$		& $-0.15$	& $38$	\\
Ba$_{3}$	& $2617$	& $8.21_{-0.01}^{+0.01}$		& $1.150_{-0.001}^{+0.006}$		& $0.093_{-0.006}^{+0.008}$		& $0.091_{-0.014}^{+0.008}$	& $-9.443_{-0.011}^{+0.004}$	& $-0.26$	& $57$	\\
\hline
Rb$_{5}$	& $869$		& $10.46_{-0.04}^{+0.02}$		& $2.2309_{-0.0081}^{+0.0008}$	& $0.575_{-0.012}^{+0.006}$		& $0.60_{-0.03}^{+0.02}$	& $-11.91_{-0.02}^{+0.04}$		& $0.18$	& $29$	\\
Ra$_{5}$	& $920$		& $9.19_{-0.08}^{+0.08}$		& $1.86_{-0.07}^{+0.04}$		& $0.262_{-0.007}^{+0.013}$		& $0.181_{-0.007}^{+0.010}$	& $-11.3_{-0.1}^{+0.2}$			& $0.29$ 	& $29$	\\
Bc$_{5}$	& $1312$	& $9.77_{-0.06}^{+0.04}$		& $1.61_{-0.03}^{+0.03}$		& $0.102_{-0.011}^{+0.005}$		& $0.656_{-0.014}^{+0.007}$	& $-10.03_{-0.03}^{+0.04}$		& $-0.15$	& $37$	\\
Bb$_{5}$	& $2744$	& $8.64_{-0.03}^{+0.03}$		& $1.196_{-0.007}^{+0.006}$		& $-0.011_{-0.007}^{+0.007}$	& $0.42_{-0.01}^{+0.01}$	& $-9.497_{-0.009}^{+0.010}$	& $-0.22$	& $45$	\\
Ba$_{5}$	& $1493$	& $8.12_{-0.03}^{+0.02}$		& $1.14_{-0.02}^{+0.01}$		& $0.16_{-0.02}^{+0.02}$		& $-0.038_{-0.005}^{+0.006}$& $-9.41_{-0.03}^{+0.03}$		& $-0.28$	& $61$	\\
\hline
Rb$_{6}$	& $771$		& $10.50_{-0.05}^{+0.01}$		& $2.236_{-0.008}^{+0.011}$		& $0.601_{-0.013}^{+0.005}$		& $0.629_{-0.032}^{+0.004}$	& $-11.91_{-0.11}^{+0.03}$		& $0.17$	& $29$	\\
Ra$_{6}$	& $790$		& $9.32_{-0.11}^{+0.03}$		& $1.94_{-0.05}^{+0.01}$		& $0.26_{-0.03}^{+0.02}$		& $0.201_{-0.017}^{+0.004}$	& $-11.69_{-0.03}^{+0.19}$		& $0.37$	& $27$	\\
Bd$_{6}$	& $1104$	& $9.90_{-0.06}^{+0.13}$		& $1.68_{-0.03}^{+0.07}$		& $0.11_{-0.02}^{+0.03}$		& $0.66_{-0.01}^{+0.01}$	& $-10.11_{-0.09}^{+0.03}$		& $-0.12$	& $37$	\\
Bc$_{6}$	& $952$		& $8.51_{-0.28}^{+0.07}$		& $1.30_{-0.09}^{+0.02}$		& $0.359_{-0.100}^{-0.007}$		& $0.04_{-0.10}^{+0.03}$	& $-9.67_{-0.03}^{+0.15}$		& $-0.21$	& $49$	\\
Bb$_{6}$	& $2454$	& $8.79_{-0.06}^{+0.12}$		& $1.24_{-0.03}^{+0.04}$		& $0.00_{-0.01}^{+0.03}$		& $0.47_{-0.02}^{+0.04}$	& $-9.58_{-0.05}^{+0.04}$		& $-0.20$	& $41$	\\
Ba$_{6}$	& $1267$	& $7.98_{-0.03}^{+0.15}$		& $1.059_{-0.006}^{+0.032}$		& $-0.04_{-0.03}^{+0.04}$		& $0.06_{-0.06}^{+0.12}$	& $-9.272_{-0.051}^{+0.009}$	& $-0.32$	& $65$	\\
\hline
\end{tabular}
\end{table*}

Furthermore, we find a basic structure of two ``superclusters'' (separated by the dashed black line) at all values of $k$, including the simplest partition $k = 2$. This indicates the strength of the bimodality in the structure of our sample. As $k$ increases, \texttt{k-means} favours splitting the blue supercluster apart over the red supercluster, due mostly to its spread in our features and the higher number of galaxies in the blue supercluster, in conjunction with the "uniform effect" of \texttt{k-means}.

We introduce a preliminary naming scheme for our clusters based on their correlation with colour. The scheme is intended as a quick way to identify clusters in the various comparisons and analyses we conduct in this section, rather than as a full description or explanation of cluster identities. This scheme is not to imply that colour is entirely responsible for the clustering outcomes (though it clearly plays a strong role). Cluster names consist of three parts in the format ``Xy$_{Z}$''. The first part, either ``R'' (for red) or ``B'' (for blue), corresponds to the supercluster (upper and lower respectively in figure \ref{fig:hi}) to which the cluster belongs. The second letter ranks the cluster by its mean $u-r$ colours in comparison with other clusters within the same supercluster at the same value of $k$. Rankings begin at ``a'' for the bluest cluster, and follow on alphabetically until all clusters within the supercluster are named. The third part, a number, indicates the solution (i.e. the value of $k$) to which the cluster belongs.

\texttt{k-means} clusters are defined by their centroids. Table \ref{tab:c} summarises the clusters in our best solutions ($k = 2, 3, 5,$ and $6$). The leftmost section lists the clusters and the numbers of galaxies that they contain ($N_{C}$). The middle section contains the cluster centroids as coordinates in each of the features, along with uncertainties estimated by application of the bootstrap method to our sample. The rightmost section lists the ``environment centroid'' of the clusters (i.e. the mean $\Sigma_{5}$ of the galaxies they contain; \emph{not} a feature used in the clustering), and the percentage of galaxies lost from each of the clusters due to incompleteness of environmental data. Sorting the clusters by their mean colour means they are also correlated with $SSFR$ and $M_{*}$ in all four solutions. This is consistent with our expectation that PC1 (with which these features are most strongly associated) would dictate much of the clustering. We note that \cite{SIUDEK+2018} also find a strong correlation of their galaxy classes (via an unsupervised method) with colour and star formation activity. $n$ and $R_{1/2}$ do not correlate as strongly with our clusters (when sorted by colour), particularly at higher $k$, indicating that these features only play a role in dictating clusters as the number of clusters increases. The broad correlations of these features with the clusters at lower values of $k$ are due to their correlations with our PC1 features. We also find a correlation of environmental density with the clusters (when sorted by colour), indicative of the strong role of environment in galaxy evolution.

In order to understand cluster structures, we reprise the panels of figure \ref{fig:samp}. We omit the original histograms to avoid visual clutter, especially at higher values of $k$. We use coloured histograms and coloured contour projections to show the distribution of our clusters in 1D and 2D feature spaces. The contours are drawn to enclose $75$ percent of the galaxies in each cluster. This level is chosen to strike a balance between generality and accuracy, given that clusters are best characterised by points near their centres. Cluster centroids are plotted as filled circles of the same colour. We retain the classical dividers for comparison with clusters.

\begin{figure*}
\centering
\includegraphics[width =0.9\textwidth]{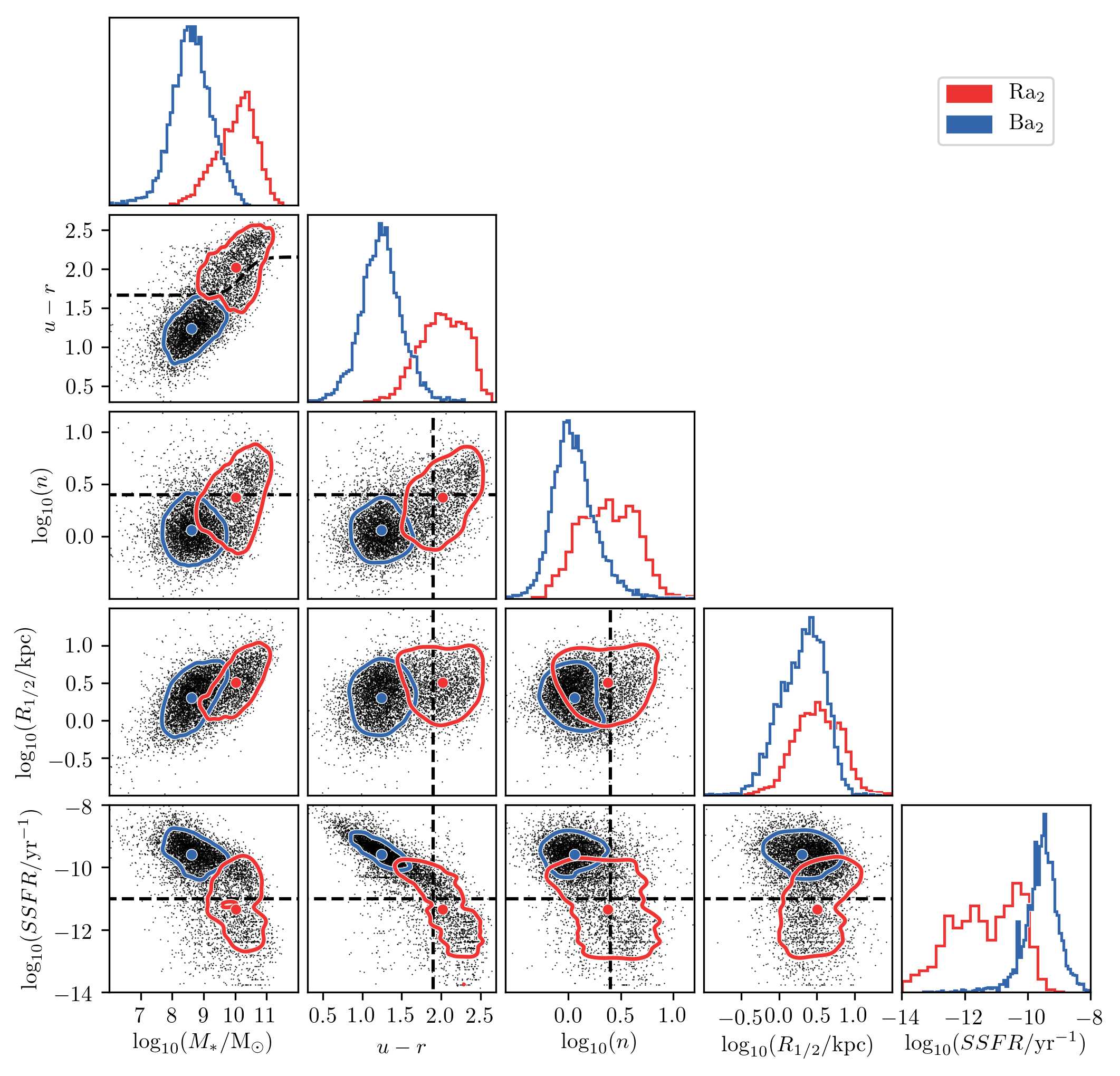}
\caption{A profile of $k = 2$. Clusters are represented using coloured histograms and contours, and their centroids are marked using filled circles of the same colour.}
\label{fig:k2}
\end{figure*}

\begin{figure*}
\centering
\includegraphics[width =\textwidth]{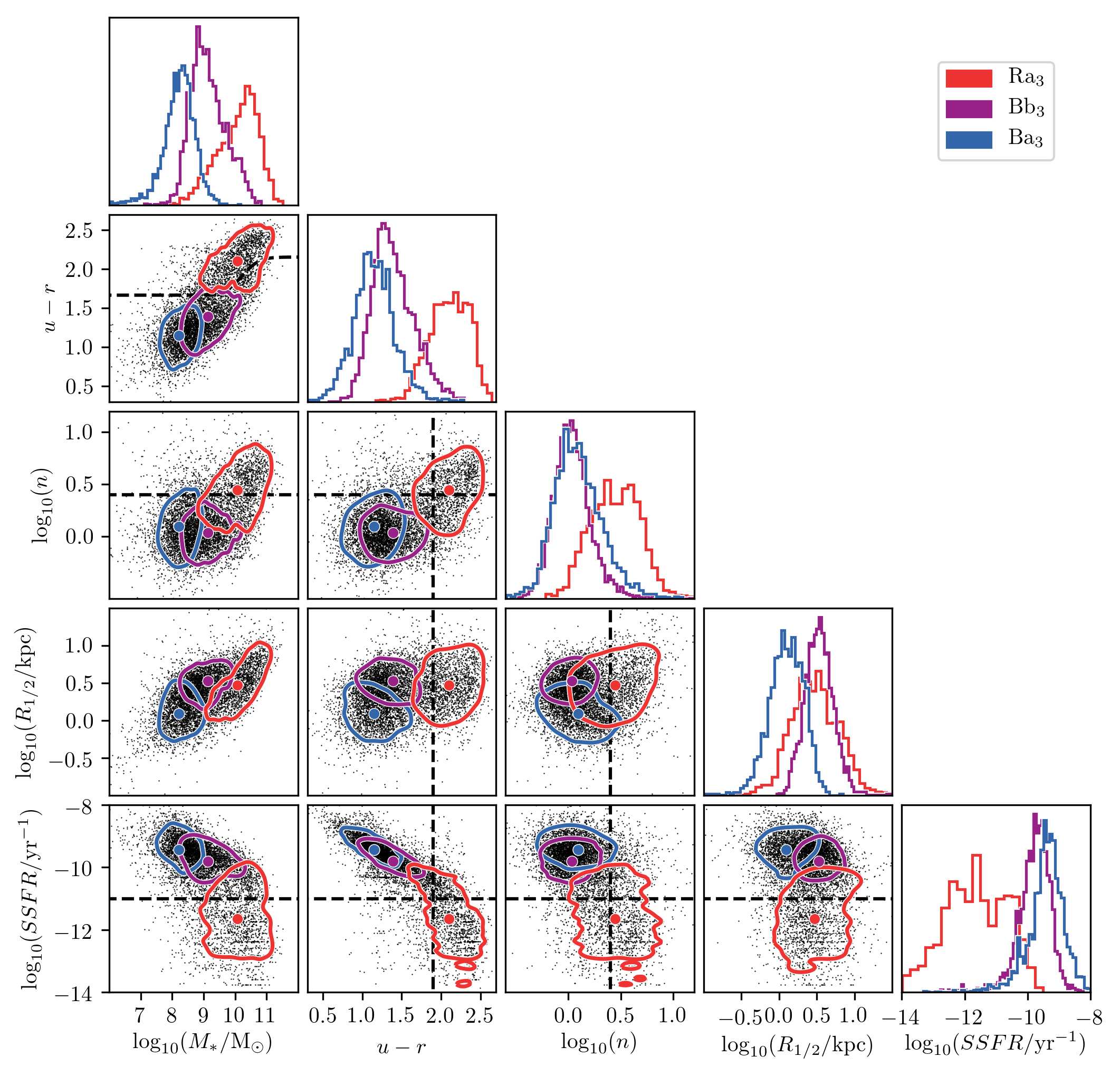}
\caption{A profile of $k = 3$. Clusters are represented using coloured histograms and contours, and their centroids are marked using filled circles of the same colour.}
\label{fig:k3}
\end{figure*}

\begin{figure*}
\centering
\includegraphics[width =0.9\textwidth]{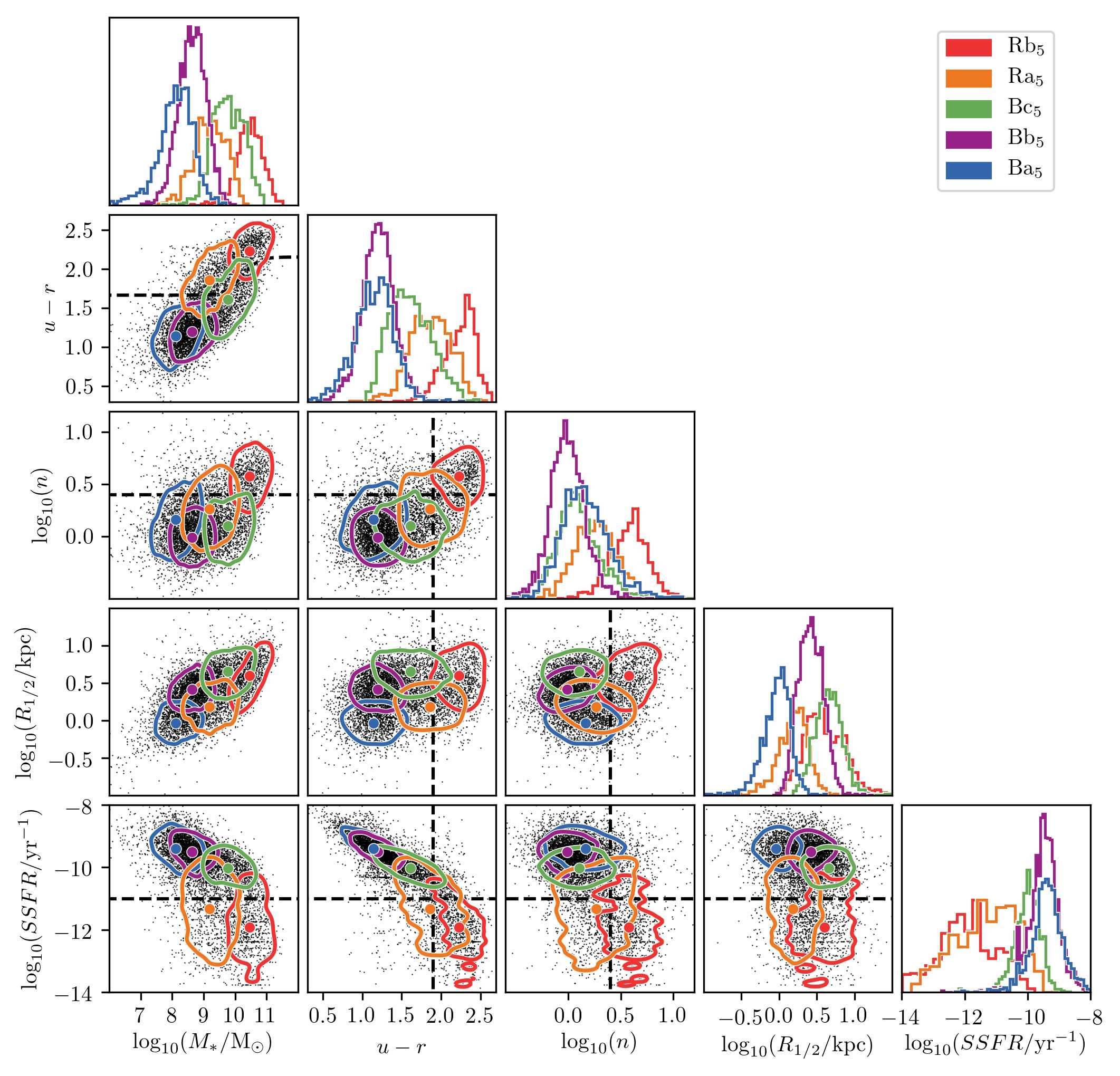}
\caption{A profile of $k = 5$. Clusters are represented using coloured histograms and contours, and their centroids are marked using filled circles of the same colour.}
\label{fig:k5}
\end{figure*}

\begin{figure*}
\centering
\includegraphics[width =\textwidth]{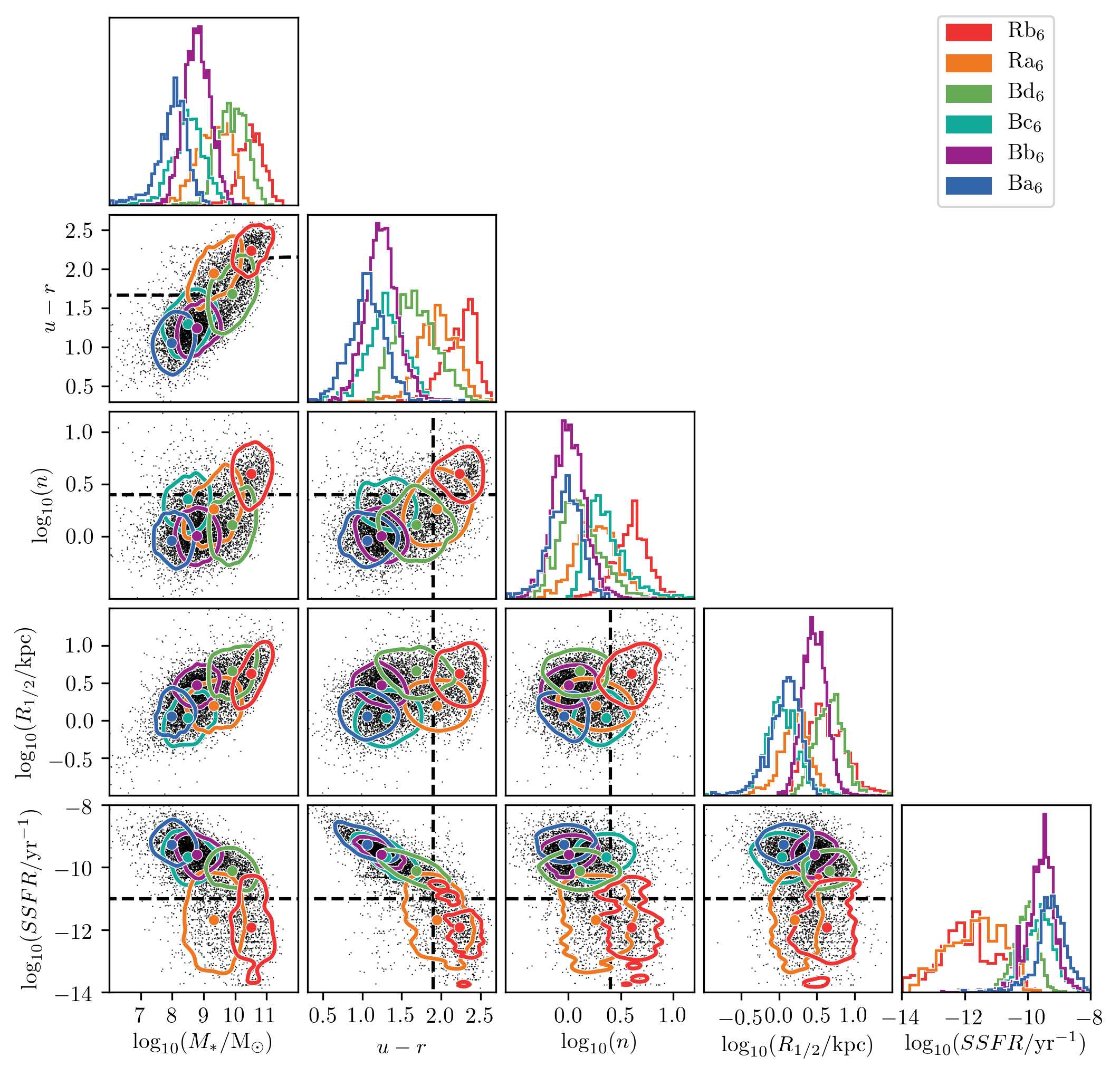}
\caption{A profile of $k = 6$. Clusters are represented using coloured histograms and contours, and their centroids are marked using filled circles of the same colour.}
\label{fig:k6}
\end{figure*}

In the following sections we describe each of the solutions in detail.

\subsection{$k = 2$}
\label{sec:k2}

\begin{figure}
\centering
\includegraphics[width=0.45\textwidth]{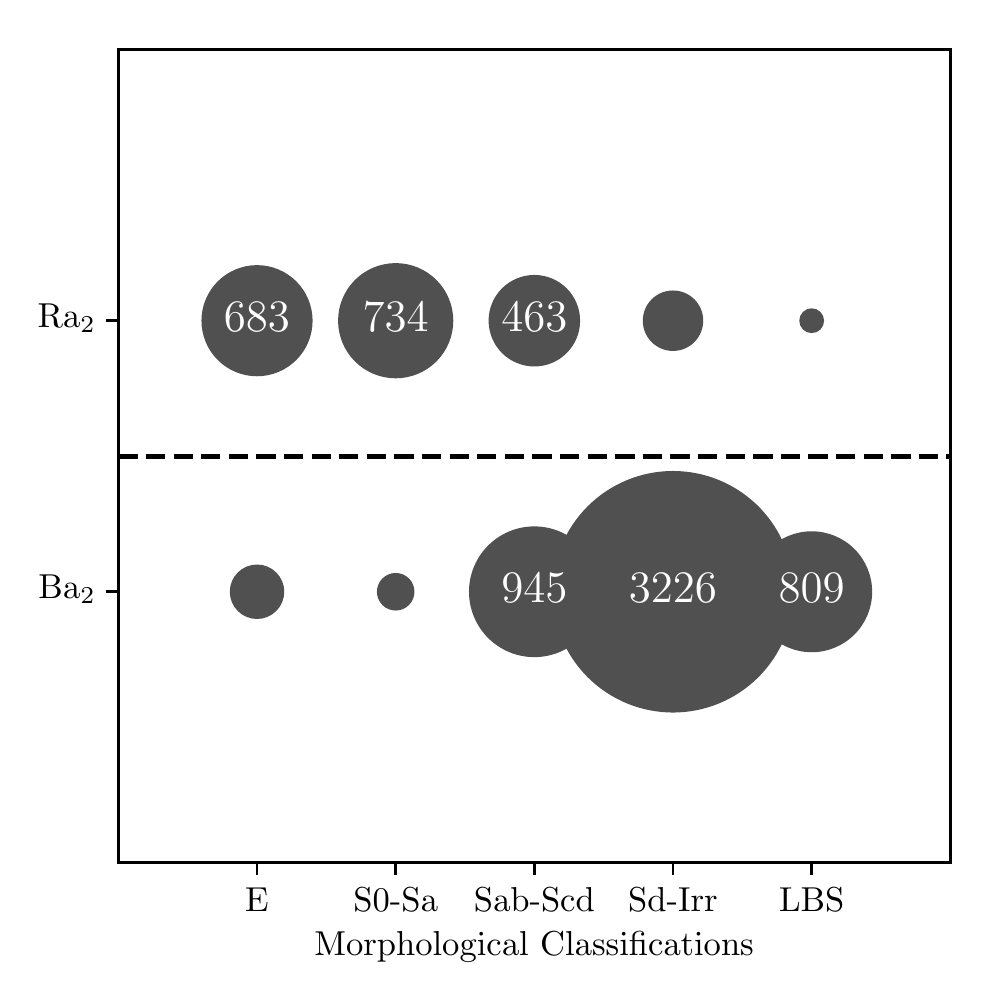}
\caption{Bubble plot comparing $k = 2$ with the \protect\cite{KELVIN+2014a} and \protect\cite{MOFFETT+2016} morphological classifications. All bubbles containing more than $5$ per cent of the galaxies in our sample are labelled with the number of galaxies that they contain. The dashed black line separates the two superclusters that \texttt{k-means} finds.}
\label{fig:h2}
\end{figure}

\begin{figure}
\centering
\includegraphics[width=0.45\textwidth]{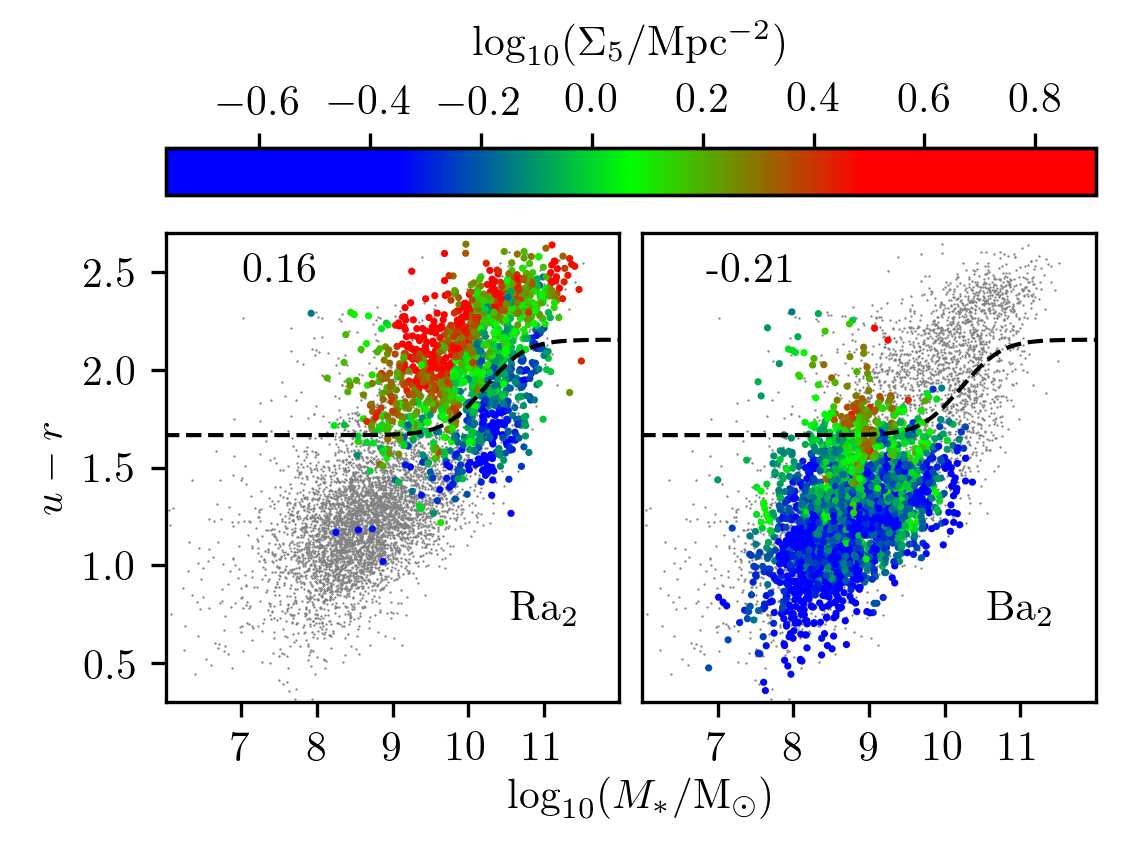}
\caption{Our $k = 2$ clusters, projected onto the $u-r$ vs. $M_{*}$ plane, with points coloured by their smoothed local environmental densities ($\Sigma_{5}$). The small grey points represent the remainder of our sample, as well as those galaxies for which $\Sigma_{5}$ is not available (including those within the clusters highlighted in each panel). Cluster names are shown in the bottom right of each panel. The mean $\Sigma_{5}$ of each cluster is shown in the top left of each panel. The dashed black line marks the \protect\cite{BALDRY+2004} distinction between the blue and red sequences of galaxies.}
\label{fig:e2}
\end{figure}

Figure \ref{fig:hi} shows that $k = 2$ forms the basic structure of two superclusters into which the clusters at higher values of $k$ may also be divided, indicating the influence of the bimodality on the clustering. Table \ref{tab:c} reveals that the clusters in $k = 2$ represent two distinct populations. Ra$_{2}$, which contains fewer galaxies than Ba$_{2}$, is made up of galaxies with higher masses, redder colours, higher S\'ersic indices, larger radii, and lower $SSFR$s on average. This is consistent with established notions of an overall bimodality of galaxies. Cluster Ra$_{2}$ has larger uncertainties on its centroid in all features in comparison with those of Ba$_{2}$ because it less dense in feature space.

The $k = 2$ cluster projections in figure \ref{fig:k2} are best separated in panels involving $u-r$, $SSFR$, and $M_{*}$. These are the features that are most strongly associated with PC1, which dominates the covariance structure of our sample in feature space and hence dictates much of the clustering. The $k = 2$ cluster projections overlap more in panels involving $R_{1/2}$ and $n$, which are more strongly associated with PC2 and PC3 respectively. These features play a lesser role in the clustering in $k = 2$. 

Cluster Ra$_{2}$ spans the classical dividers (dashed black lines) in all panels of figure \ref{fig:k2}. While appearing to represent red sequence galaxies \citep{BALDRY+2004,TAYLOR+2015} and quiescent (negligible $SSFR$) galaxies (table \ref{tab:c}), it extends well onto the blue sequence ($u-r$ vs. $M_{*}$) and star-forming main sequence ($SSFR$ vs. $M_{*}$). This is because the cluster boundary that \texttt{k-means} draws between its two centroids is a hyperplane, equidistant from them both and perpendicular to the line connecting them. The uniform effect of \texttt{k-means}, which produces clusters of similar sizes, essentially bisects our sample through PC1, along which the centroids are evenly spaced. This gives a coarse partition of our sample. More clusters are needed to properly ``resolve'' the true structure and boundary of the bimodality of galaxies.

In figure \ref{fig:h2}, we use a bubble plot to visualise agreement between our clusters and the \cite{KELVIN+2014a} and \cite{MOFFETT+2016} morphological classifications. The plot shows a considerable overlap of morphologies between the two clusters, verifying the relative weakness of $n$ and $R_{1/2}$ in the clustering in $k = 2$. This effect is also seen in figure \ref{fig:k2}, which shows that cluster Ra$_{2}$ is indiscriminate with respect to S\'ersic indices in comparison with our classical divider. The broad correlation of clusters with morphological types (e.g. that earlier type morphologies are more likely to be found in cluster Ra$_{2}$; figure \ref{fig:h2}) arises as a result of the correlation of morphology with our PC1 features. This effect is apparent in figure \ref{fig:ps2}, which shows that the galaxies in Ra$_{2}$ have smoother and more concentrated morphologies.

The mean local environmental densities (table \ref{tab:c}) of the $k = 2$ clusters reveal that the galaxies in Ra$_{2}$ occupy denser environments on average than those in Ba$_{2}$. We note that this is mostly a reflection of the basic correlation of galaxy mass, colour, and star formation activity with environment, due to the coarse partition of our sample. A greater fraction of galaxies are lost from Ba$_{2}$ than Ra$_{2}$, such that this difference is likely to be underestimated. The panels in figure \ref{fig:e2} project the cluster onto the $u-r$ vs. $M_{*}$ plane. Points are coloured by their smoothed local environmental densities ($\Sigma_{5}$; see section \ref{sec:dat}). Ba$_{2}$ consists mostly of galaxies in low density environments. We note a gradual increase of $\Sigma_{5}$ with $u-r$ within Ba$_{2}$. Ra$_{2}$ exhibits more of a spread in $\Sigma_{5}$, but with a preference for higher density environments. This suggests that environmental processes are the more common mechanism by which galaxies acquire redder colours.

\subsection{$k = 3$}
\label{sec:k3}

\begin{figure}
\centering
\includegraphics[width=0.45\textwidth]{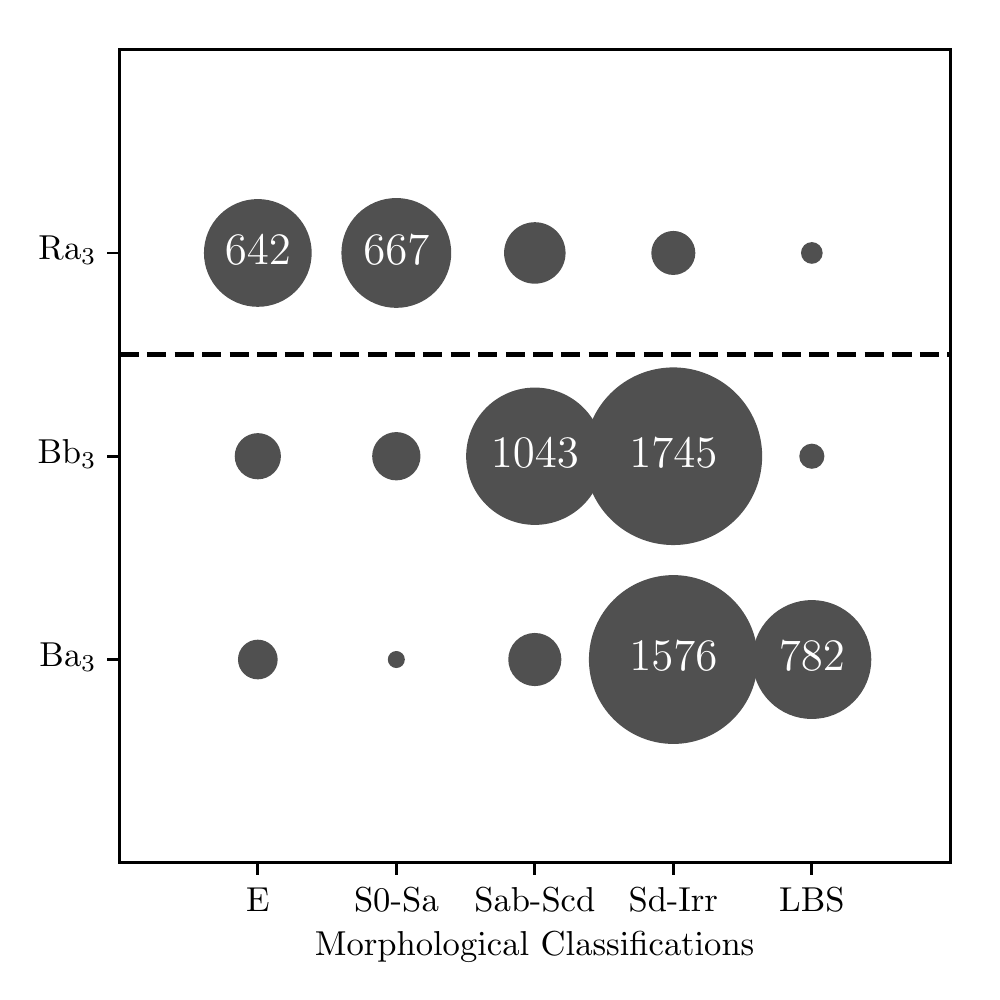}
\caption{Bubble plot comparing $k = 3$ with the \protect\cite{KELVIN+2014a} and \protect\cite{MOFFETT+2016} morphological classifications. All bubbles containing more than $5$ per cent of the galaxies in our sample are labelled with the number of galaxies that they contain. The dashed black line separates the two superclusters that \texttt{k-means} finds.}
\label{fig:h3}
\end{figure}

\begin{figure}
\centering
\includegraphics[width=0.45\textwidth]{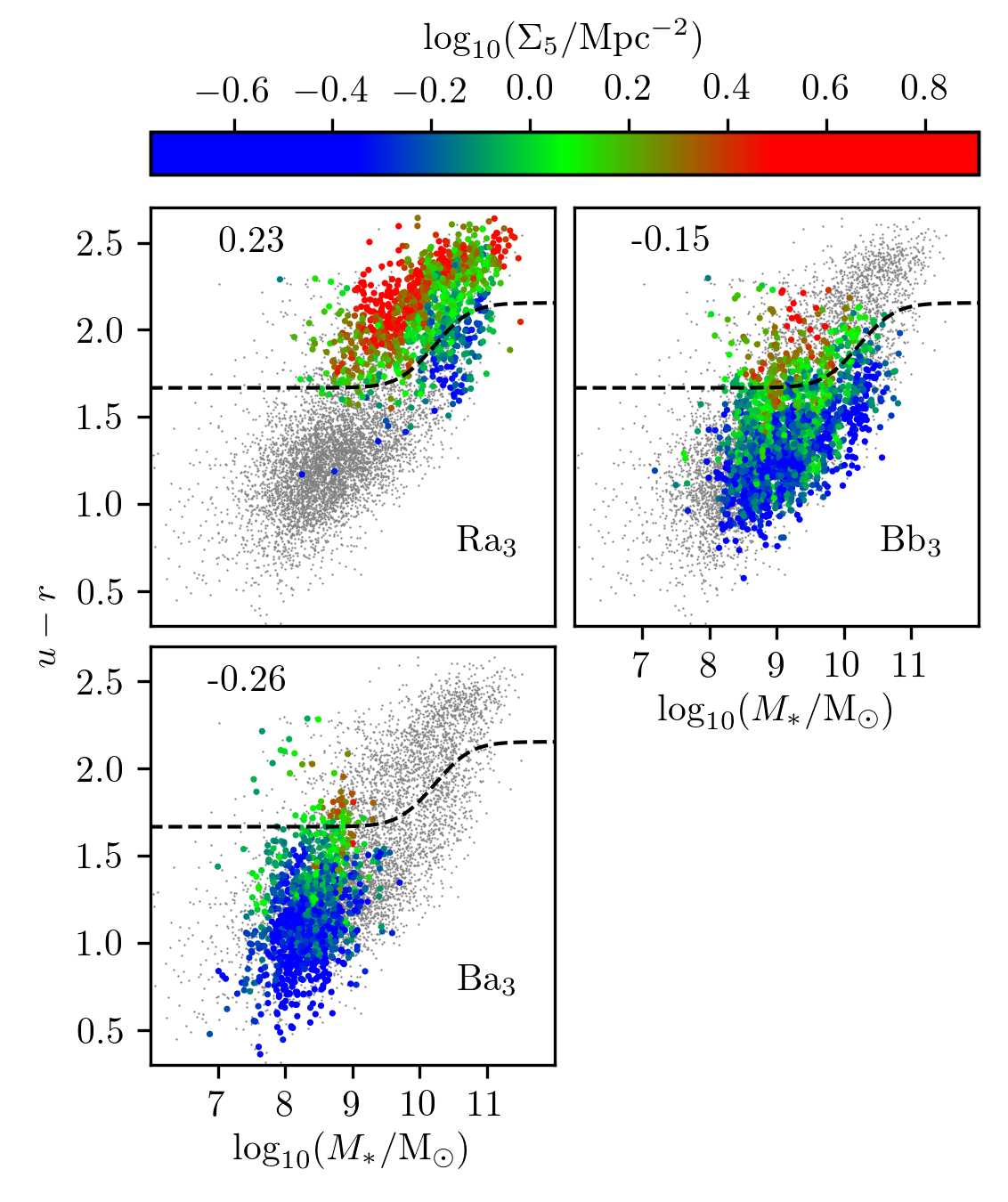}
\caption{Our $k = 3$ clusters, projected onto the $u-r$ vs. $M_{*}$ plane, with points coloured by their smoothed local environmental densities ($\Sigma_{5}$). The small grey points represent the remainder of our sample, as well as those galaxies for which $\Sigma_{5}$ is not available (including those within the clusters highlighted in each panel). Cluster names are shown in the bottom right of each panel. The mean $\Sigma_{5}$ of each cluster is shown in the top left of each panel. The dashed black line marks the \protect\cite{BALDRY+2004} distinction between the blue and red sequences of galaxies.}
\label{fig:e3}
\end{figure}

Figure \ref{fig:hi} shows that $k = 3$ is hierarchical with respect to $k = 2$; only $\sim 6$ per cent of the galaxies in our sample do not follow a clean hierarchy between the two solutions. Therefore, much of the cluster structure of $k = 3$ is derived from that of $k = 2$. The red supercluster remains relatively unchanged between the two solutions. Ra$_{3}$ contains $\sim 76$ per cent of galaxies that Ra$_{2}$ contains, and table \ref{tab:c} shows that both clusters share similar identities. The placement of Ra$_{3}$ with respect to the classical dividers is improved in most panels, particularly those involving our PC1 features ($M_{*}$, $u-r$, and $SSFR$), indicating the strength with which they still dictate the clustering in $k = 3$. This improvement is enabled by the split of the blue supercluster, which has evened out the cluster sizes. The uncertainties on the centroid of Ra$_{3}$ are generally less than or equal to those of Ba$_{3}$ and Bb$_{3}$ except for in $SSFR$, which is due to the spread in $SSFR$ of the quiescent galaxies in Ra$_{3}$ (see figure \ref{fig:k3}).

The main change in $k = 3$ from $k = 2$ is that \texttt{k-means} splits the blue supercluster apart into two clusters: Ba$_{3}$ and Bb$_{3}$. The split happens due to the larger number of galaxies within the blue supercluster. The main features that distinguish these clusters are $R_{1/2}$ and $M_{*}$, in which the centroids (table \ref{tab:c}) differ by $1.42\sigma$ and $1.03\sigma$ respectively (as opposed to the $< 0.50\sigma$ differences in other features). Here, $\sigma$ is the standard deviation of our sample in a given feature. Figure \ref{fig:k3} shows that the distributions of the clusters are most distinct in these features as well. Ba$_{3}$ and Bb$_{3}$ exhibit very similar distributions in $n$ and $SSFR$. While differing in mass and size, galaxies in both Ba$_{3}$ and Bb$_{3}$ are generally star-forming and have diffuse light profiles.

The use of an additional centroid has enabled \texttt{k-means} to explore the subtler variance in our sample (and in particular, in the blue supercluster) that PC2 encompasses. The particularly low masses and sizes of the galaxies in Ba$_{3}$ suggest a distinction by \texttt{k-means} between dwarf galaxies and larger, more massive galaxies in the blue supercluster. Figure \ref{fig:h3} appears to confirm this, showing that the more evolved spiral galaxies in our sample are more likely to be found in cluster Bb$_{3}$. Figure \ref{fig:ps3} shows that the galaxies in Bb$_{3}$ have more prominent disks. The clusters still exhibit significant overlap in morphologies and cluster Ra$_{3}$ is still indiscriminate with respect to $n$, indicating a continuing relative weakness of $n$ in dictating the clustering at $k = 3$.

The galaxies in Ra$_{3}$ occupy denser environments on average than those in either Ba$_{3}$ or Bb$_{3}$ (see table \ref{tab:c}). Figure \ref{fig:e3} shows that clusters Ba$_{3}$ and Bb$_{3}$ are similarly distributed in $\Sigma_{5}$. Both are dominated by low density environments and exhibit tails towards higher density environments. For Bb$_{3}$ we note that galaxies in low density environments are found at all masses, suggesting that some galaxies are able to evolve to higher masses without any significant change in their morphology (see figure \ref{fig:h3}) due to, for example, major mergers. The dwarf-like galaxies in Ba$_{3}$, on the other hand, are more likely, as satellites, to be affected by their environments during infall, such that they do not evolve to higher masses in low density environments.

\subsection{$k = 5$}
\label{sec:k5}

\begin{figure}
\centering
\includegraphics[width=0.45\textwidth]{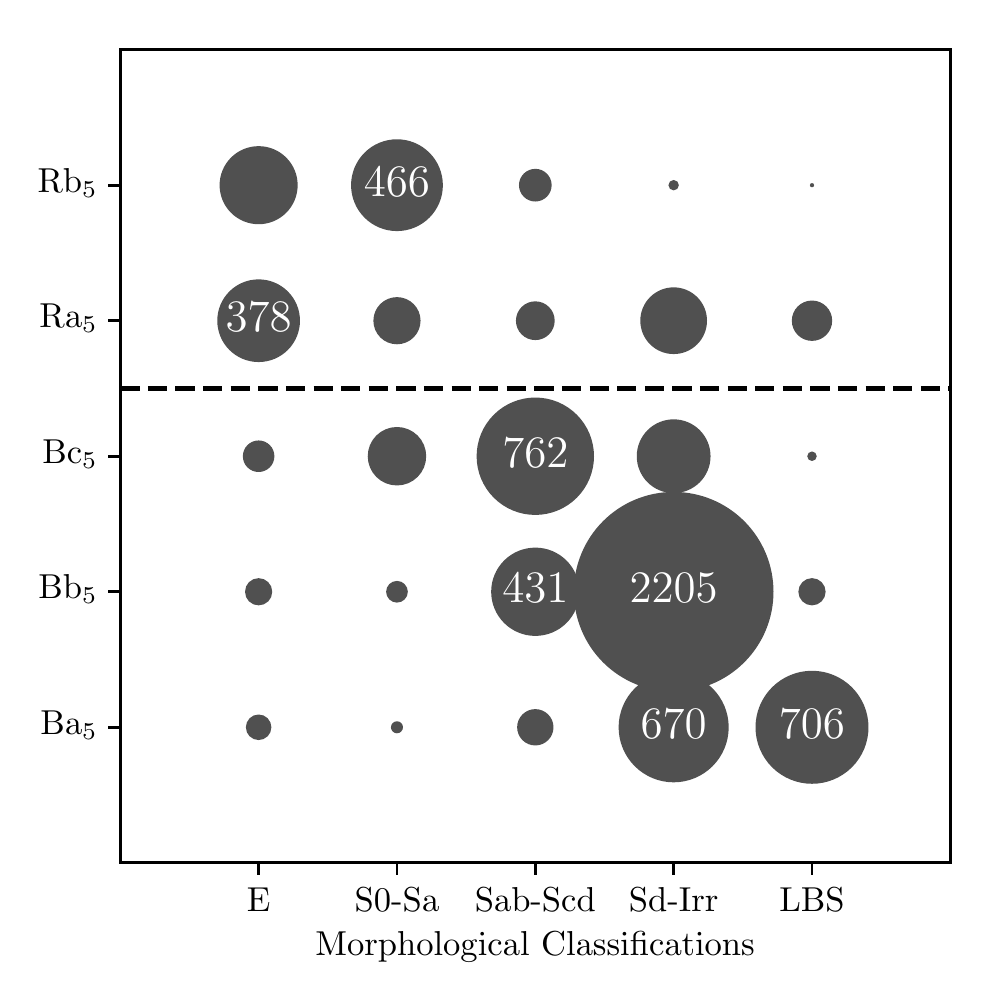}
\caption{Bubble plot comparing $k = 5$ with the \protect\cite{KELVIN+2014a} and \protect\cite{MOFFETT+2016} morphological classifications. All bubbles containing more than $5$ per cent of the galaxies in our sample are labelled with the number of galaxies that they contain. The dashed black line separates the two superclusters that \texttt{k-means} finds.}
\label{fig:h5}
\end{figure}

\begin{figure}
\centering
\includegraphics[width=0.45\textwidth]{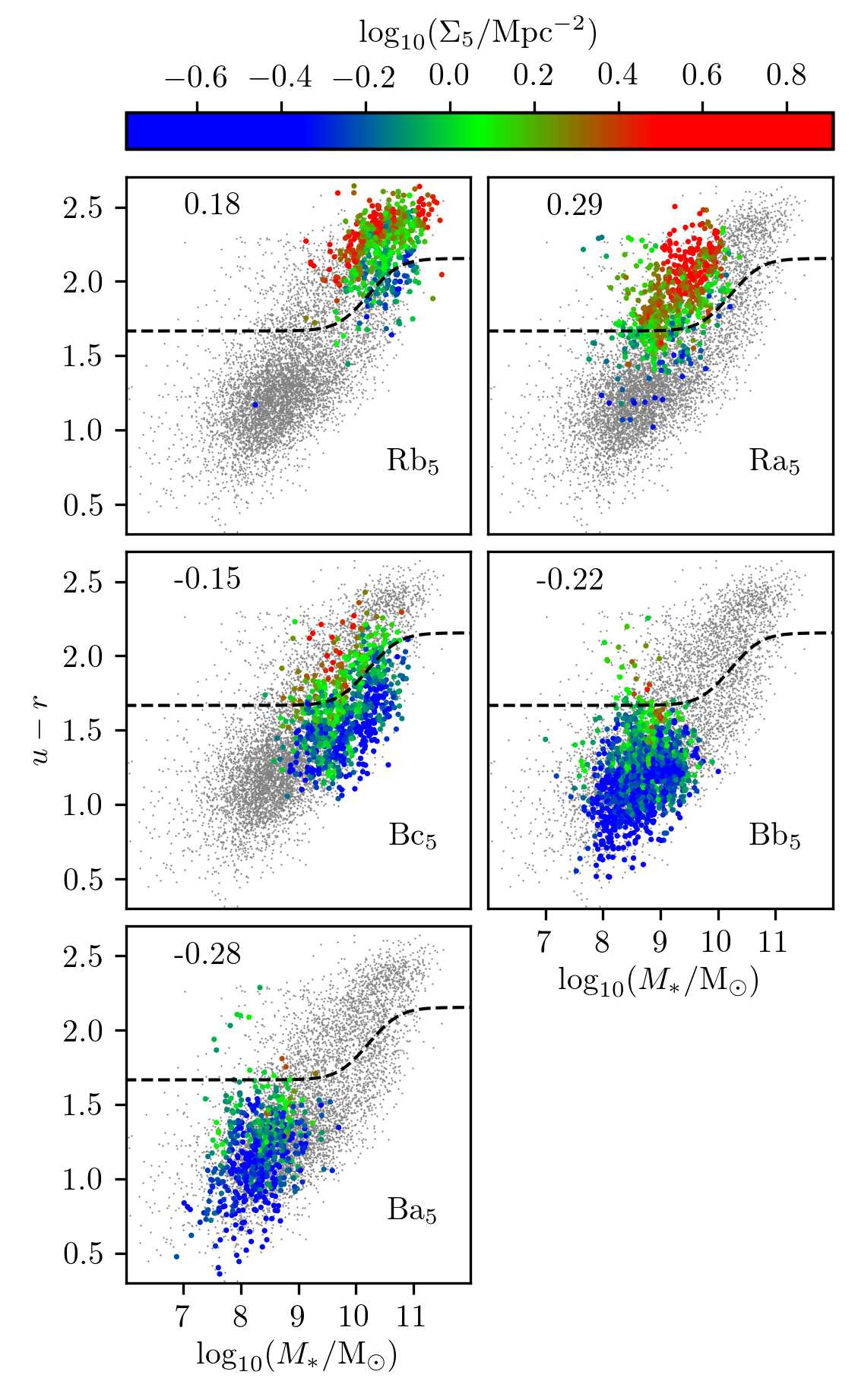}
\caption{Our $k = 5$ clusters, projected onto the $u-r$ vs. $M_{*}$ plane, with points coloured by their smoothed local environmental densities ($\Sigma_{5}$). The small grey points represent the remainder of our sample, as well as those galaxies for which $\Sigma_{5}$ is not available (including those within the clusters highlighted in each panel). Cluster names are shown in the bottom right of each panel. The mean $\Sigma_{5}$ of each cluster is shown in the top left of each panel. The dashed black line marks the \protect\cite{BALDRY+2004} distinction between the blue and red sequences of galaxies.}
\label{fig:e5}
\end{figure}

$k = 5$ is not as cleanly hierarchical with respect to $k = 3$ as $k = 3$ is with respect to $k = 2$ (figure \ref{fig:hi}). There is mixing between the red and blue superclusters and within the blue supercluster, which involves $\sim 21$ per cent of the galaxies in our sample. This indicates that \texttt{k-means} has probed an alternative structure of our sample in feature space to that which it finds in $k = 2$ and $k = 3$. Our exclusion of a solution from $k = 4$ due to instability (figure \ref{fig:stab}) may exaggerate the apparent mixing. The bimodal stability of solutions at $k = 4$ emerges as \texttt{k-means} settles into a split in \emph{either} the red or the blue supercluster, rather than splits in both as in $k = 5$. Both of these splits must be made in order to achieve stability, suggesting genuine differences between the galaxies occupying these clusters.

The red supercluster is split into two clusters: Ra$_{5}$ and Rb$_{5}$. The features that distinguish these clusters are $M_{*}$, $R_{1/2}$, and $n$, with differences of $1.41\sigma$, $1.35\sigma$, and $1.19\sigma$ respectively in their centroids (see table \ref{tab:c}). We note that they are also separated in the $SSFR$ vs. $u-r$ plane. Rb$_{5}$ consists mostly of evolved galaxies with the highest masses and reddest colours in our sample. Enabled by the use of additional centroids to probe subtler variances in the sample, \texttt{k-means} has also now made a morphological distinction between galaxies such that Rb$_{5}$ contains galaxies with the highest S\'ersic indices as well. This is also apparent in figures \ref{fig:h5} and \ref{fig:ps5}, which respectively show that Rb$_{5}$ is made up mostly of early type galaxies and of concentrated, smooth, spheroid-dominated galaxies.

Ra$_{5}$ has a weaker cluster identity. While Rb$_{5}$ contains galaxies from only one cluster in $k = 3$, Ra$_{5}$ contains galaxies from three. Its centroid has larger uncertainties in most features than those of the other $k = 5$ clusters (table \ref{tab:c}), and it exhibits a large spread in figure \ref{fig:k5}, spanning the red and blue sequences and including both star-forming and quiescent galaxies. The galaxies Ra$_{5}$ contains have red colours and low specific star formation rates like those in Rb$_{5}$, but not to the same extent, suggesting that they are not as evolved. Figure \ref{fig:h5} reveals that it also contains a range of morphologies, with even E and Sd-Irr type galaxies being grouped together. This suggests a lack of morphological information in our feature selection, despite the role $n$ is now playing in dictating cluster Rb$_{5}$. We also note an apparent inability of \texttt{k-means} to properly distinguish E and S0-Sa type galaxies. \cite{NOLTE+2018} find a similar inseparability of these morphological classes in the same sample using both supervised and unsupervised methods.

The blue supercluster is split again in $k = 5$. The mixing between $k = 3$ and $k = 5$ within the blue supercluster means the clusters in each solution do not correspond as strongly to one another. The galaxies in Ba$_{5}$ and Bb$_{5}$ have been distinguished from those in Bc$_{5}$ by their masses, in which their centroids both differ from that of Bc$_{5}$ by $> 1\sigma$. Ba$_{5}$ and Bb$_{5}$ are similar in terms of the colours and specific star formation rates of the galaxies they contain. The main distinction between them is in their sizes ($> 1\sigma$). Figure \ref{fig:ps5} shows that Bb$_{5}$ contains more galaxies with extended disks, while Ba$_{5}$ contains more galaxies that are compact. Ba$_{5}$ contains the vast majority of LBS galaxies in our sample (figure \ref{fig:h5}), while Bb$_{5}$ contains a significant number of more evolved spiral galaxies.

Bc$_{5}$ differs from the other two clusters in the blue supercluster, containing relatively massive and large galaxies with reduced star formation. The intermediate colours of the galaxies in Bc$_{5}$, and its location in the colour-mass plane in particular are consistent with previous descriptions of green valley galaxies \citep{SALIM+2007, SCHAWINSKI+2014}. The low S\'ersic indices of the galaxies in Bc$_{5}$ in comparison with those of the galaxies in the red supercluster are due to the presence of prominent disks (see figures \ref{fig:h5} and \ref{fig:ps5}). Bc$_{5}$ contains spiral galaxies at the later stages of their evolution.

Just as the use of additional centroids has enabled \texttt{k-means} to make finer distinctions between galaxies in terms of the features we use to describe them, it has also led to a finer view of the role of environment in influencing our clusters. The galaxies in clusters Ba$_{5}$, Bb$_{5}$, Bc$_{5}$ mostly occupy low density environments (table \ref{tab:c} and figure \ref{fig:e5}). Meanwhile, clusters Ra$_{5}$ and Rb$_{5}$ generally contain galaxies in higher density environments. Notably, the average local environmental densities of galaxies in clusters Bc$_{5}$ and Ra$_{5}$ differ, despite that these clusters are adjacent in all panels of figure \ref{fig:k5} and span both the blue and red sequences in figure \ref{fig:e5}.

We suggest that these ``green valley'' clusters, Ra$_{5}$ and Bc$_{5}$, each mostly contain galaxies on different evolutionary pathways. The evolution of galaxies in Ra$_{5}$, whose distribution in $\Sigma_{5}$ is skewed toward higher densities, is dominated by external processes (i.e. environment quenching; \citealt{PENG+2010}), which both transform their morphologies and inhibit their star formation on short timescales. Examples of external processes include major \citep{BARNES-1992}, and minor mergers \citep{TOOMRE+1972}. These processes are likely responsible for the early type morphologies (figures \ref{fig:h5} and \ref{fig:ps5}), red colours, and inhibited star formation rates (figure \ref{fig:k5}) in Ra$_{5}$.

Cluster Bc$_{5}$, which is dominated by galaxies in lower density environments, contains galaxies that are dominated in their evolution by internal processes. Internal processes include mass quenching, in which feedback from stars \citep{GEACH+2014} and AGN \citep{CROTON+2006} scales with galaxy stellar mass and drives star forming gas out of galaxies, and morphological quenching \citep{FANG+2013}, in which bulges at the centre of late-type galaxies stabilise their disks against collapse and thereby prevent further star formation. The high masses (figure \ref{fig:k5}) and prominent bulges (figure \ref{fig:ps5}) of the galaxies in Bc$_{5}$ seem to confirm the dominance of these internal processes in their evolution.

The differences in the morphologies of the galaxies in Ra$_{5}$ and Bc$_{5}$ is consistent with \cite{SCHAWINSKI+2014}, who find a morphological dichotomy of galaxies in the green valley. The spread in morphologies in Ra$_{5}$ may arise due to both its large spread in $\Sigma_{5}$, and the short timescales of morphological transformations. The dominance of galaxies with high $\Sigma_{5}$ in Rb$_{5}$ suggests a preference of external processes for moving galaxies onto the red sequence over time. The additional presence of galaxies in low density environments in Rb$_{5}$ suggests that galaxies evolving mostly via internal processes will also converge on the red sequence, though.

\subsection{$k = 6$}
\label{sec:k6}

\begin{figure}
\centering
\includegraphics[width=0.45\textwidth]{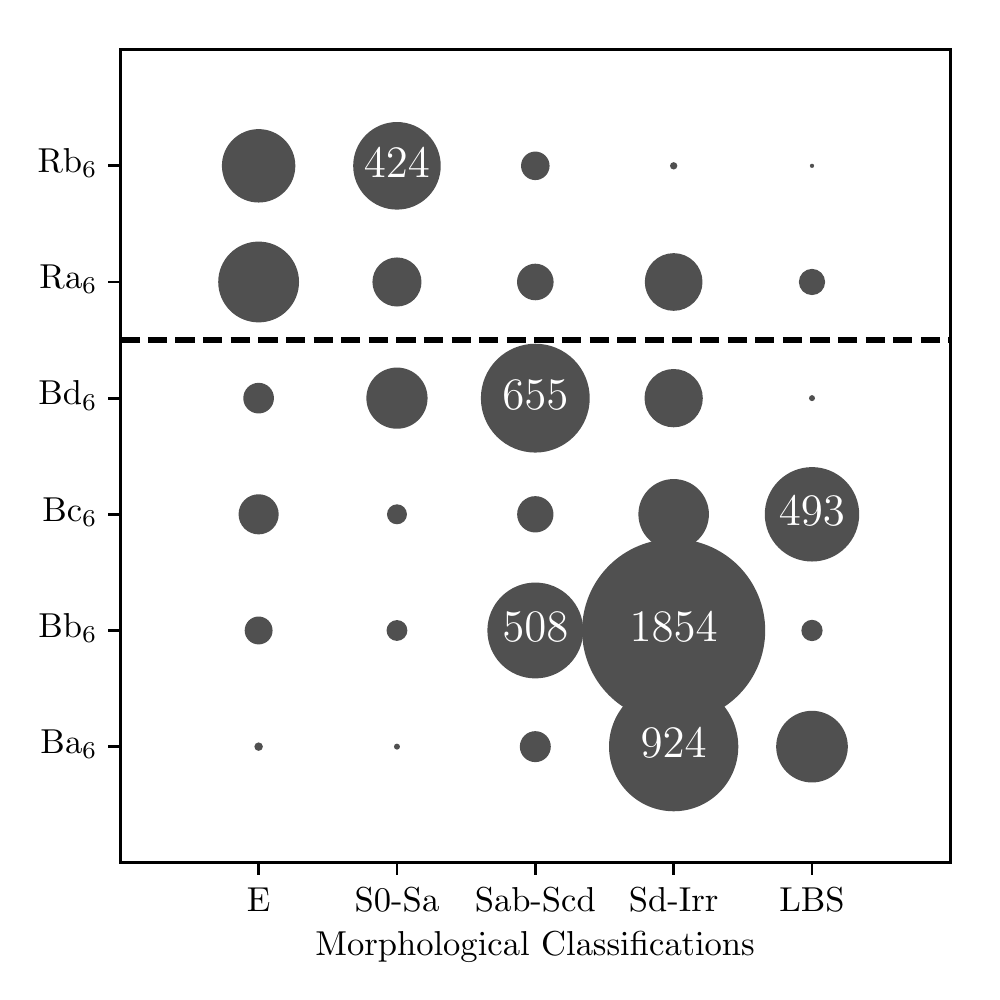}
\caption{Bubble plot comparing $k = 6$ with the \protect\cite{KELVIN+2014a} and \protect\cite{MOFFETT+2016} morphological classifications. All bubbles containing more than $5$ per cent of the galaxies in our sample are labelled with the number of galaxies that they contain. The dashed black line separates the two superclusters that \texttt{k-means} finds.}
\label{fig:h6}
\end{figure}

\begin{figure}
\centering
\includegraphics[width=0.45\textwidth]{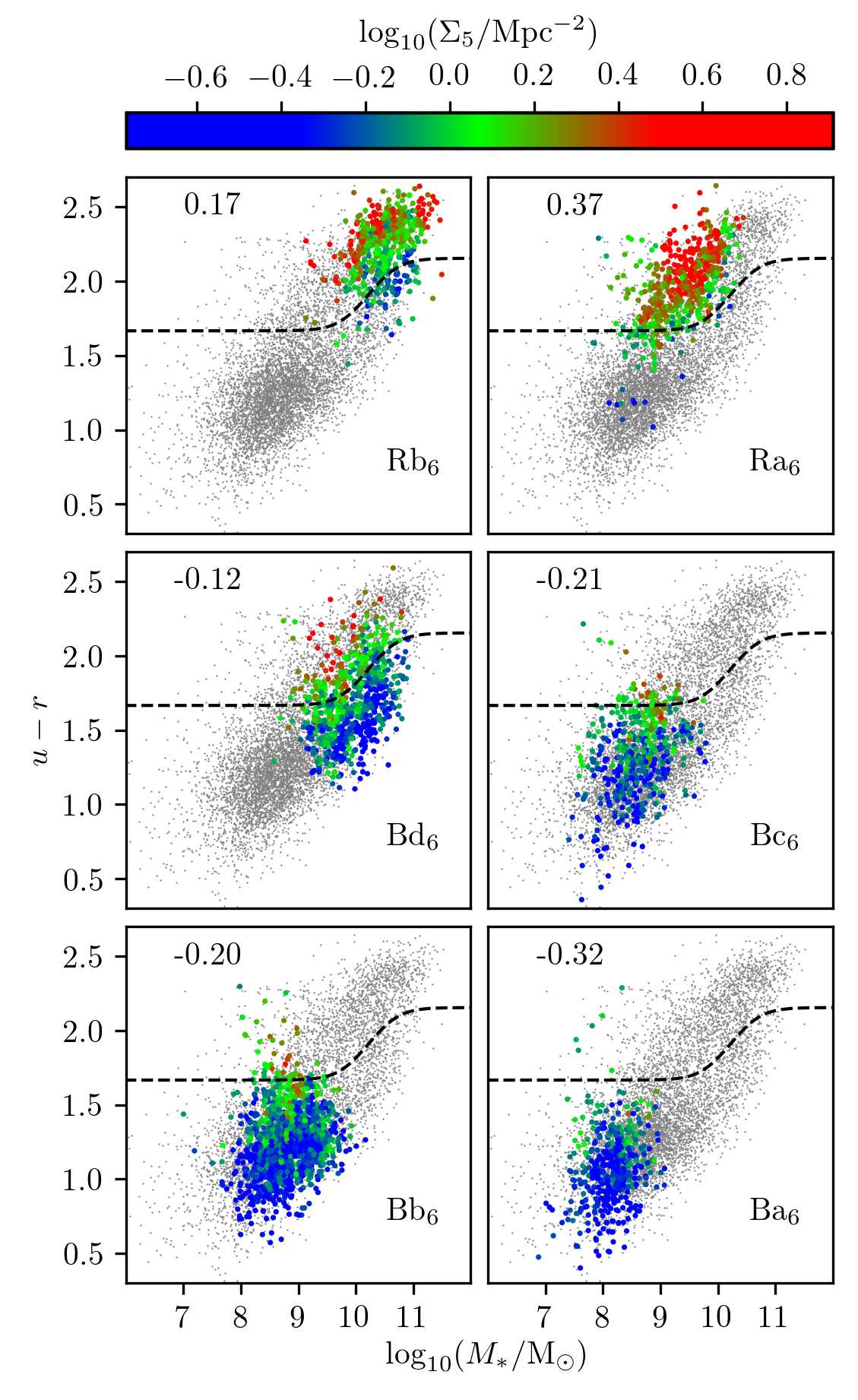}
\caption{Our $k = 6$ clusters, projected onto the $u-r$ vs. $M_{*}$ plane, with points coloured by their smoothed local environmental densities ($\Sigma_{5}$). The small grey points represent the remainder of our sample, as well as those galaxies for which $\Sigma_{5}$ is not available (including those within the clusters highlighted in each panel). Cluster names are shown in the bottom right of each panel. The mean $\Sigma_{5}$ of each cluster is shown in the top left of each panel. The dashed black line marks the \protect\cite{BALDRY+2004} distinction between the blue and red sequences of galaxies.}
\label{fig:e6}
\end{figure}

$k = 6$ is once again more strongly hierarchical with respect to $k = 5$ than $k = 5$ is with respect to $k = 3$; $\sim 15$ per cent of galaxies mix between $k = 5$ and $k = 6$. Clusters are more readily comparable with those in $k = 5$, from which most of their structure is derived. The clusters in the red supercluster retain their identities in terms of their centroids (table \ref{tab:c}), distributions (figure \ref{fig:k6}), and morphologies (figure \ref{fig:h6}) between $k = 5$ and $k = 6$ and remain mostly unchanged. Similarly, Bd$_{6}$ matches well with Bc$_{5}$, as does Bb$_{6}$ with Bb$_{5}$.

The main changes between $k = 5$ and $k = 6$ are at the blue end of the blue supercluster. Ba$_{6}$ and Bc$_{6}$ both have low masses and small radii. They differ most significantly in $n$ ($1.34\sigma$), appearing to indicate a morphological distinction between disk-dominated galaxies in Ba$_{6}$ and spheroid-dominated galaxies in Bc$_{6}$. While figure \ref{fig:h6} reveals a significant degeneracy of morphological classifications in these clusters, the distinction is more apparent in figure \ref{fig:ps6}. The classification degeneracy may arise partially due to the difficulty incurred in visually classifying intrinsically faint objects. We note that Bc$_{6}$ contains some potentially spurious spheroid-dominated E type galaxies, which could explain some of the difference in $n$ between Ba$_{6}$ and Bc$_{6}$. Further morphological information in our feature selection may lead to a clearer distinction between galaxies at the blue end of the blue supercluster.

Clusters Ba$_{6}$, Bb$_{6}$, and Bd$_{6}$, which all contain disky galaxies (figures \ref{fig:h6} and \ref{fig:ps6}) in mostly low density environments (figure \ref{fig:e6}), appear to form a continuum of galaxy evolution along the blue sequence. This continuum is dictated by internal processes, given the increase in mass and bulge prominence along the sequence of consistently low $\Sigma_{5}$. The environments of the galaxies in Bc$_{6}$ are also low density, though they have an early type morphology, suggesting that they may have formed differently. We note a significant tail of this cluster toward intermediate densities, suggesting that they may be in the early stages of morphological transformation due to environmental effects. Their origin and fate is unclear \citep{SCHAWINSKI+2009}.

We note that agreement of our clusters with the classical dividers has improved considerably as $k$ has increased. At $k = 6$, the clusters align particularly well with established notions of a fundamental bimodality of galaxies, which we express using the dashed black lines in figure \ref{fig:k6}. Cluster Ra$_{6}$ still spans the dividers in some panels, though this may be due to the rapid timescales of the morphological transformations that the galaxies it contains are likely undergoing, such that they exhibit a larger spread in our features.

\section{Summary}
\label{sec:summ}

We report the results of a test of the $k$-means unsupervised clustering method as a galaxy classification solution for the unprecedentedly large surveys of the future and as a tool for exploring feature spaces of high dimensionalities. It is tested on a redshift- and magnitude-limited sample of $7338$ galaxies from the GAMA survey, which we represent using a preliminary selection of five features: stellar mass, $u-r$ colour, S\'ersic index, half-light radius, and specific star formation rate. Analyses of correlations and covariances between features reveal that stellar mass, $u-r$ colour, and specific star formation rate dominate much of the structure of our sample in feature space and hence dictate much of the clustering. We rescale, truncate, and normalise our sample ahead of clustering to mitigate a) the influence of outliers on results, and b) bias toward any of the features based on skewed distributions or large numerical ranges. 

We apply the $k$-means method, which partitions data into $k$ clusters, in the context of a unique cluster evaluation approach that enables us to robustly identify stable clustering structure in spite of stochastic effects, including the initialisation of \texttt{k-means}, and application of the bootstrap method to our sample. We note that to examine the stability of clustering in our sample at a single value of $k$ takes just $\sim 3$ minutes using a single core on a laptop computer. This framework is therefore highly scalable. We find that the local galaxy population is stably divisible into $2,3,5,$ and a maximum of $6$ clusters. We select optimal clustering solutions from each value of $k$ for analysis. We reach the following conclusions:

\begin{enumerate}
\setcounter{enumi}{0}

\item Clusters in all four of our best solutions agree with established notions of the bimodality of galaxies. Agreement improves as $k$ increases. The use of additional centroids to model the data structure of our sample in feature space enables a more detailed view of the bimodality via \texttt{k-means}. At higher values of $k$, we find distinct clusters that appear to follow different evolutionary pathways through the green valley. While $M_{*}$, $u-r$, and $SSFR$ dictate most of the clustering structure in all four of our best solutions, $n$ and $R_{1/2}$ play an increasingly strong role at higher $k$ as \texttt{k-means} uses the additional centroids to explore subtler variances in our sample.

\item Though we do not aim to reproduce any existing classification schemes with our clusters, we find a general agreement of our clusters with the \cite{KELVIN+2014a} and \cite{MOFFETT+2016} Hubble-like morphological classifications of the galaxies in our sample. At low $k$, this agreement is mostly due to the correlation of morphology with stellar mass, $u-r$ colour, and specific star formation rate, which dictate the majority of the clustering. This suggests a relative lack of morphological information among our feature selection. At higher $k$, though, we find that \texttt{k-means} is able to explore subtler variances in our sample and make genuine morphological distinctions between galaxies using S\'ersic index. The addition of further morphological features to our selection is anticipated to improve these distinctions.

\item Analysis of the local environmental densities of the galaxies in the clusters in solutions $k = 5$ and $k = 6$ especially suggests the differential roles of internal and external processes in galaxy evolution. Those clusters containing more galaxies in high density environments also contain more galaxies with early type morphologies, with their spreads in morphologies indicating rapid morphological transformation and their reduced $SSFRs$ indicating quenching. Clusters containing galaxies in low density environments are found along the whole blue sequence, such that some galaxies are able to evolve to the highest masses while retaining a disky morphology. Clusters in the blue sequence appear to form an evolutionary continuum of clusters whose galaxies are dominated in their evolution by internal processes. There is an apparent preference for externally driven evolution of low mass ($M_{*} < 10^{10}$ M$_{\odot}$) galaxies onto the red sequence. An availability of this environmental data for the entirety of our sample would significantly strengthen these (or alternative) conclusions regarding the role of environment in the evolution of galaxies.

\end{enumerate}

We endorse $k = 6$ as being the most useful solution for its ability to both capture the broad bimodal structure of the galaxy population in feature space and identify finer distinctions \emph{within} this bimodality that highlight the differential role of environment in the evolution of galaxies.

While we caution against the use of too many features due to issues such as overfitting and redundancy, it is clear that our feature selection may be improved by the addition of further information. We suggest that the inclusion of morphological features like asymmetry or distance-independent smoothness \citep{CONSELICE-2003}, the Gini coefficient \citep{ABRAHAM+2003}, or those derived from two-component fits might yield stronger cluster identities and further disentangle the roles of internal and external evolutionary processes, particularly with respect to clusters Ra$_{5}$ and Ra$_{6}$ which both contain a spread of morphologies. \cite{WIJESINGHE+2010} show that morphological information from multiple bands of photometry also incorporates information about the stellar populations within galaxies. We anticipate that the inclusion of spectroscopic features would also improve clustering results and cluster interpretation. In particular, emission line ratios such as those used in BPT diagrams \citep{BALDWIN+1981} could highlight the role of active galactic nuclei in galaxy evolution, and the strength of the 4000 Angstrom break \cite{POGGIANTI+1997} could include galaxy ages in cluster identities. In general, it appears that an optimal feature selection will consist of a combination of features derived from both photometry and spectroscopy. 

Our feature selection for the work in this paper is preliminary, and based mostly astrophysical knowledge. While some simple statistical consideration is applied (Spearman rank-order correlations and a principal component analysis), a variety of other methods are also available (see chapter 2 of \citealt{AGGARWAL+2014} for a comprehensive review) for feature selection and feature extraction (i.e. the manufacture of features) which may further improve clustering results.

Our sample is well-characterised by a number of previous studies, facilitating the interpretation of the clusters that we find, but it is small and limited to low redshifts. While our sample suffices for an initial test of \texttt{k-means} and our cluster evaluation approach, a more thorough test would be to apply the framework to a larger sample of galaxies spanning a greater range of redshifts, constituting a more complete representation of the diversity of galaxies in and beyond the local Universe. The Sloan Digital Sky Survey would be particularly suitable given the wealth of features available, and especially given its overlap with the Galaxy Zoo project which would enable more detailed study of morphologies within clusters. Furthermore, clustering a sample of galaxies spanning a greater range of redshifts, or a comparison of clusters in different redshift bins, invites the examination of the evolution of clusters themselves over cosmic time.

We conclude by emphasising that our cluster evaluation approach is malleable. It may readily be adapted for use with any algorithm and any sample to identify stable clustering structure. We test stability against random initialisations and application of the bootstrap method to our sample, but the approach may also be applied in the context of other Monte Carlo methods. 

Our $k = 2,3,5,$ and $6$ classifications are available in a catalogue at \url{http://www.astro.ljmu.ac.uk/~aststurn/mstar_ur_n_hlr_ssfr_tlz_k2356.txt}. A Python 3 script containing functions that implement the $k$-means method and our cluster evaluation approach is available at \url{https://github.com/sebturner/stacopy}.

\section*{Acknowledgements}

The work in this paper made use of the scikit-learn \citep{PEDREGOSA+2011}, matplotlib \citep{HUNTER-2007}, scipy \citep{JONES+2001}, and numpy \citep{OLIPHANT-2006} packages for Python 3, and the Starlink Tool for OPerations on Catalogues And Tables (TOPCAT; \citealt{TAYLOR-2005}).

GAMA is a joint European-Australasian project based around a spectroscopic campaign using the Anglo-Australian Telescope. The GAMA input catalogue is based on data taken from the Sloan Digital Sky Survey and the UKIRT Infrared Deep Sky Survey. Complementary imaging of the GAMA regions is being obtained by a number of independent survey programmes including GALEX MIS, VST KiDS, VISTA VIKING, WISE, Herschel-ATLAS, GMRT and ASKAP providing UV to radio coverage. GAMA is funded by the STFC (UK), the ARC (Australia), the AAO, and the participating institutions. The GAMA website is \url{http://www.gama-survey.org/}.

ST is funded via a Science and Technology Facilities Council postgraduate studentship. We thank Malgorzata Siudek and Steve Phillipps for helpful comments.

\bibliographystyle{mnras}
\bibliography{bib}

\appendix

\section{Stability Simulation}
\label{app:sim}

\begin{figure}
\centering
\includegraphics[width=0.23\textwidth]{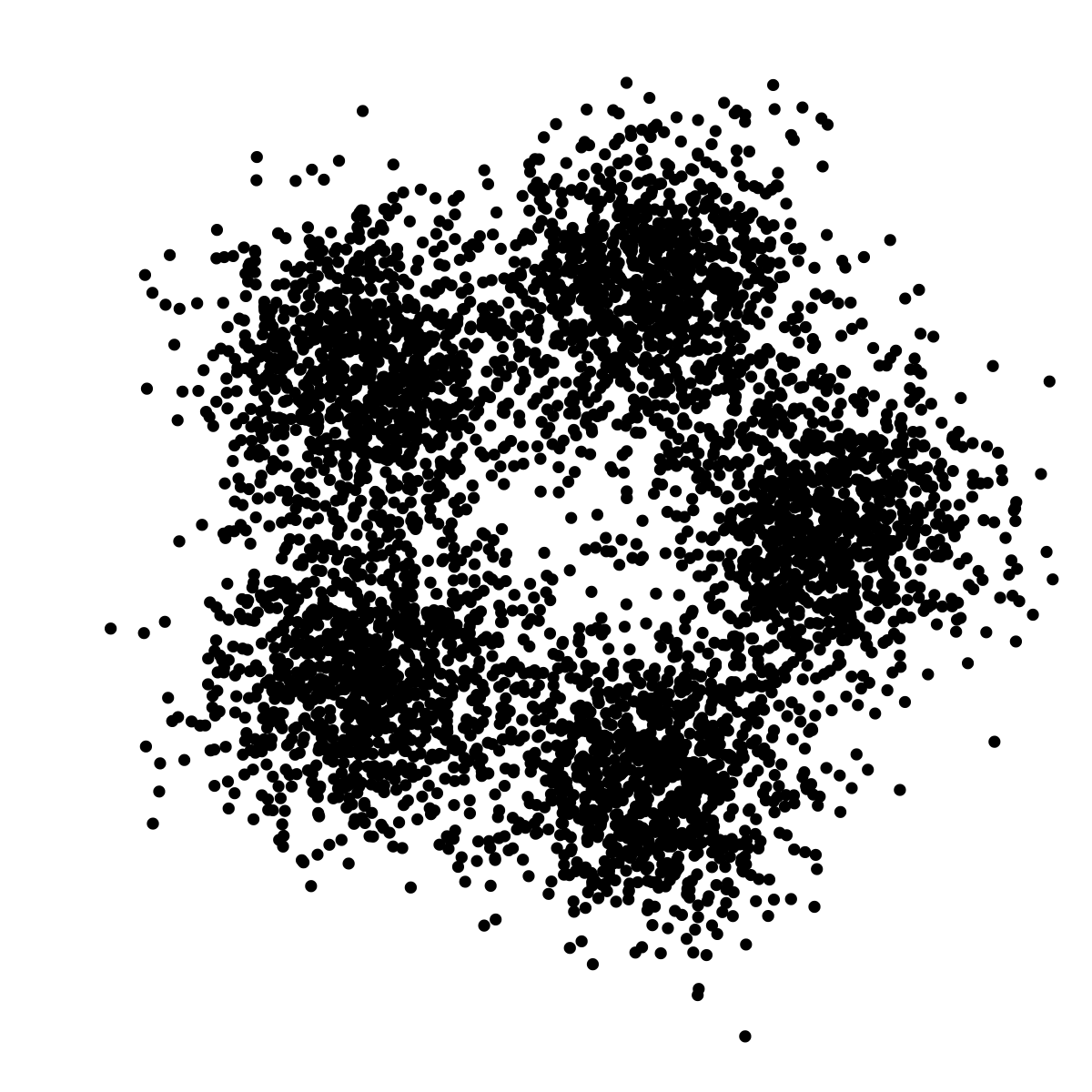}
\centering
\includegraphics[width=0.23\textwidth]{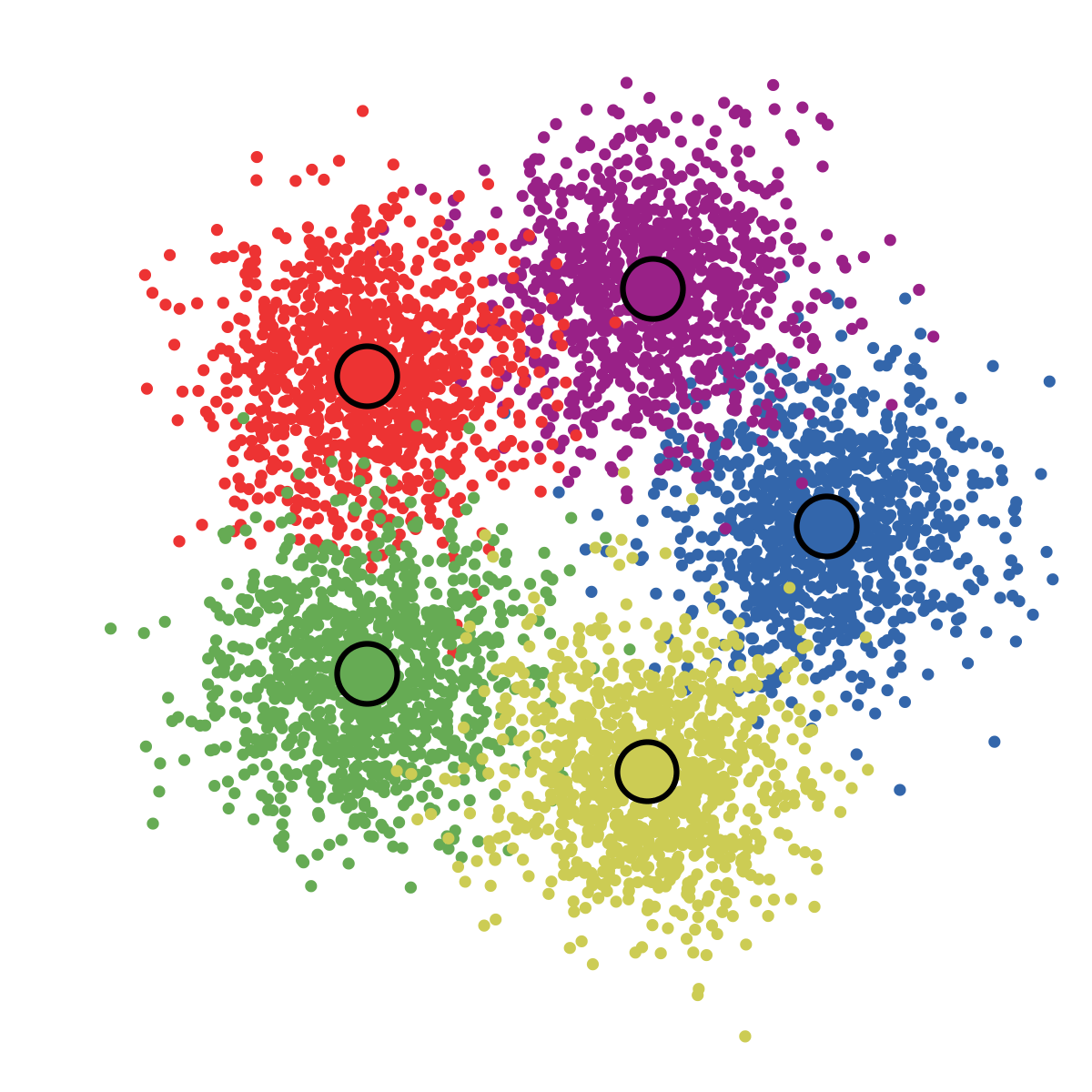}
\caption{On the left-hand side we display our simulation, containing five true clusters. See the main text for information on how it is generated. On the right-hand side we colour the points by their truth labels; all points with the same colour belong to the same true cluster, whose centroid is marked out by a large filled circle, also of the same colour.}
\label{fig:sim}
\end{figure}

To demonstrate the use of stability for selecting good values of $k$, we set up a simple simulation (figure \ref{fig:sim}). 5000 data points are distributed equally over five 2D Gaussian functions, centred at the vertices of a unit regular pentagon. The standard deviations of the distributions ($\sigma=0.3$) are set such that they overlap slightly. The value of $k_{true}$ for this simulation is $5$. We run \texttt{k-means} with $k = 4, 5,$ and $6$. We initialise 200 times at each $k$ using the \cite{ARTHUR+2007} technique. We introduce a naming scheme for the clusters that \texttt{k-means} finds in our simulation. Cluster names consist of three parts in the format ``X$Y_{Z}$''. The first part, either ``A'' or ``B'', corresponds to a particular solution to which the cluster belongs, and is used to identify solutions in the figures in this appendix. The second part, a number, corresponds to the individual cluster, also shown in the figures. The third part, another number, indicates the value of $k$ at which the solution was found.

\begin{figure}
\centering
\includegraphics[width=0.23\textwidth]{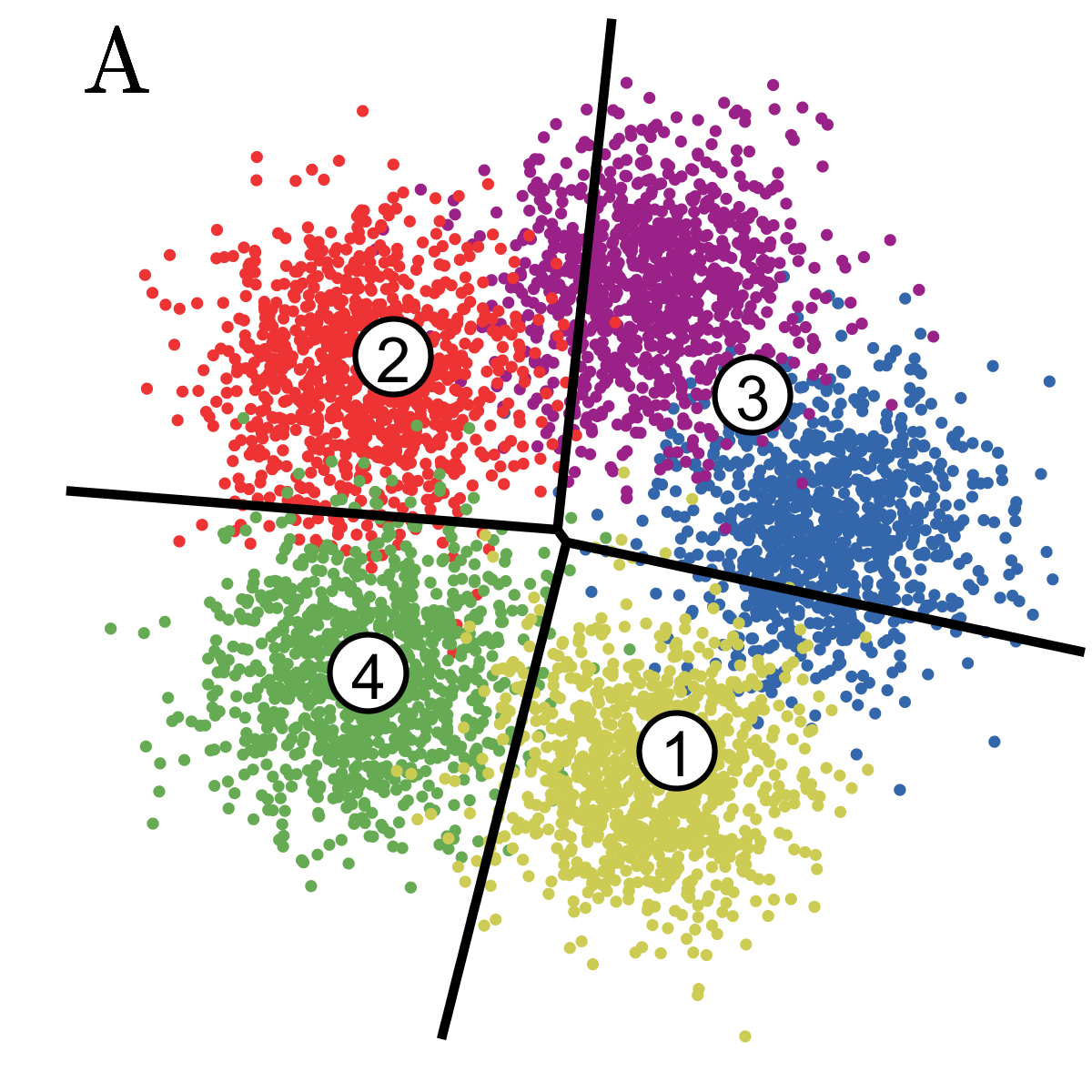}
\includegraphics[width=0.23\textwidth]{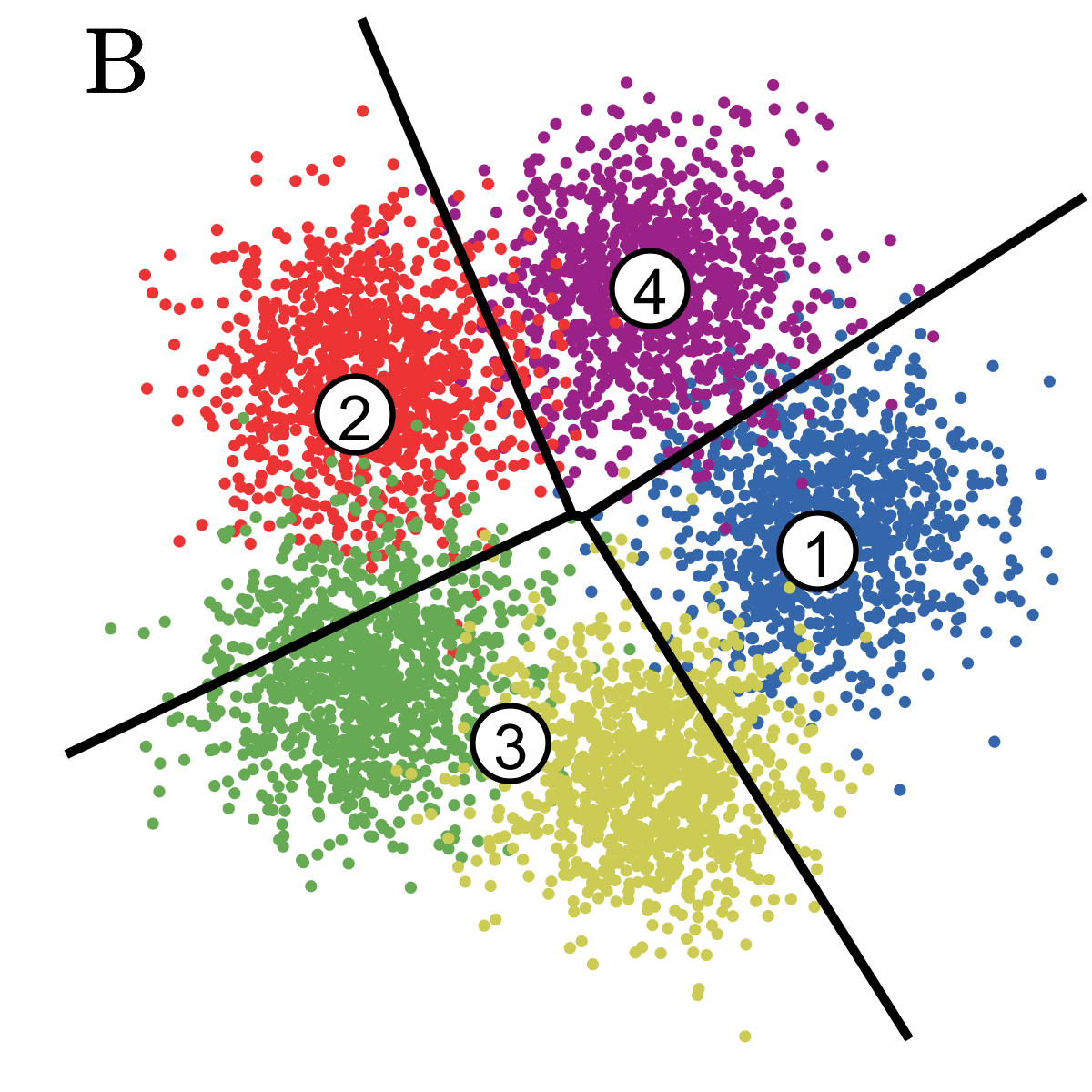}
\caption{Examples of \texttt{k-means} clustering at $k = 4 < k_{true}$. The algorithm has merged the purple and blue true clusters in solution A on the left, and the yellow and green true clusters in solution B on the right. The \texttt{k-means} centres are marked by filled white circles. The boundaries between \texttt{k-means} clusters are marked by straight black lines.}
\label{fig:sim4}
\end{figure}

Figure \ref{fig:sim4} shows two examples of the solutions found at $k = 4 < k_{true}$. In both cases, \texttt{k-means} merges two true clusters: purple and blue in solution A on the left, and yellow and green in solution B on the right. These mergers have affected the accuracy of the neighbouring \texttt{k-means} clusters as well in that they suffer from contamination (in terms of the true cluster structure). Table \ref{tab:sim4}, a contingency table (a.k.a. cross tabulation), shows that the solutions are only weakly associated with one another. The chi-squared value (equation \ref{eq:chi}) for these two solutions (A and B), calculated using the contingency table, is $6617.95$. From this, using equation \ref{eq:v}, we calculate $V = 0.66$ (with $N = 5000$ and $k = 4$).

\begin{table}
\caption{Contingency table, comparing solutions A and B generated at $k = 4$ and shown in figure \ref{fig:sim4}.}
\label{tab:sim4}
\centering
\begin{tabular}{l | r r r r}
\hline
Cluster		& B1$_{4}$	& B2$_{4}$	& B3$_{4}$	& B4$_{4}$	\\
\hline
A1$_{4}$	& $330$		& $0$		& $830$		& $0$		\\
A2$_{4}$	& $0$		& $946$		& $0$		& $194$		\\
A3$_{4}$	& $808$		& $0$		& $1$		& $839$		\\
A4$_{4}$	& $0$		& $245$		& $807$		& $0$		\\
\hline
\end{tabular}
\end{table}

\begin{figure}
\centering
\includegraphics[width=0.23\textwidth]{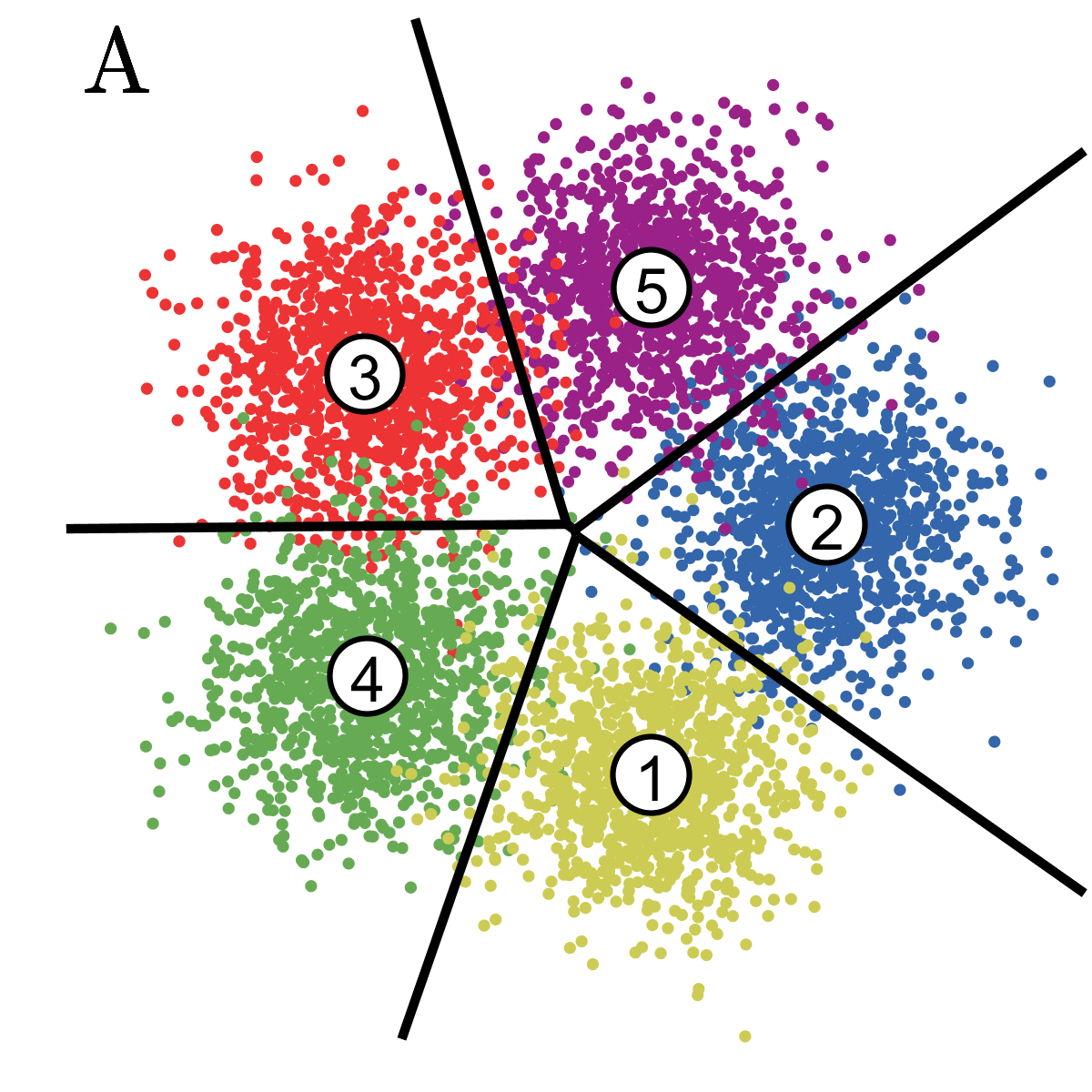}
\includegraphics[width=0.23\textwidth]{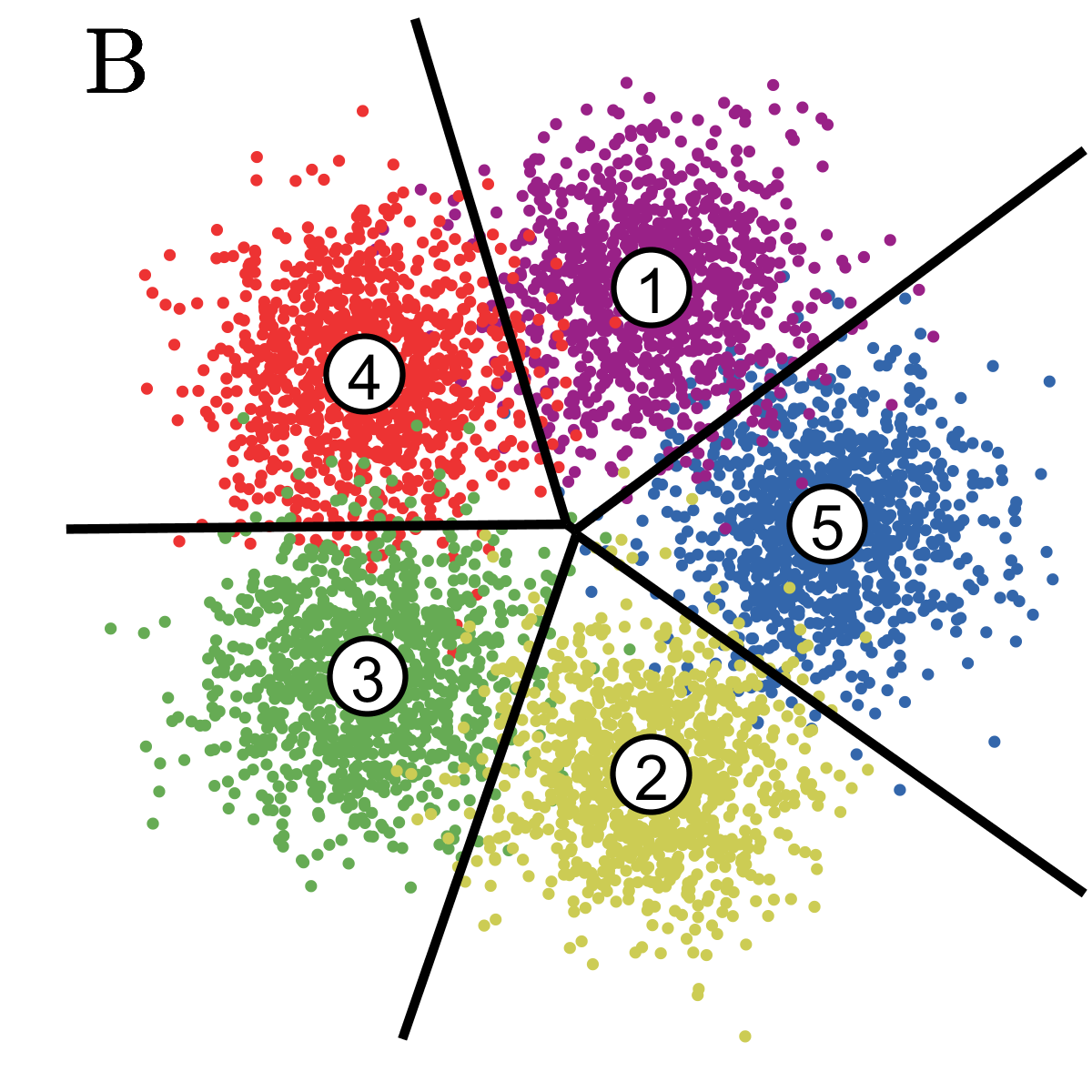}
\caption{Examples of \texttt{k-means} clustering at $k = k_{true} = 5$. The algorithm has correctly found the five true clusters in both solutions A and B, which differ by only $4$ points. The $k$-means centres are marked by filled white circles. The boundaries between \texttt{k-means} clusters are marked by straight black lines.}
\label{fig:sim5}
\end{figure}

Figure \ref{fig:sim5} shows two examples of the solutions found when $k = 5 = k_{true}$. While they appear identical, they actually differ by four points (see contingency table \ref{tab:sim4}, which shows the near-perfect association between the two solutions). \texttt{k-means} has succeeded in finding the five true clusters in both solutions. While is not impossible that \texttt{k-means} might find an alternative structure in the simulation at $k = 5$ given more initialisations, the rate at which it would do so would be so low (less than at most $0.5$ per cent given figure \ref{fig:sim5}) that $k = 5$ would still stand out as being particularly stable. For these two $k = 5$ solutions (A and B), we calculate $\chi^{2} = 19960.03$ and (with $N = 5000, k = 5$) $V = 0.999$.

\begin{table}
\caption{Contingency table, comparing solutions A and B generated at $k = 5$ and shown in figure \ref{fig:sim5}.}
\label{tab:sim5}
\centering
\begin{tabular}{l | r r r r r}
\hline
Cluster		& B1$_{5}$	& B2$_{5}$	& B3$_{5}$	& B4$_{5}$	& B5$_{5}$	\\
\hline
A1$_{5}$	& $0$		& $985$		& $0$		& $0$		& $0$		\\
A2$_{5}$	& $0$		& $3$		& $0$		& $0$		& $1005$	\\
A3$_{5}$	& $0$		& $0$		& $0$		& $993$		& $0$		\\
A4$_{5}$	& $0$		& $0$		& $1013$	& $1$		& $0$		\\
A5$_{5}$	& $1000$	 & $0$		& $0$		& $0$		& $0$		\\
\hline
\end{tabular}
\end{table}

\begin{figure}
\centering
\includegraphics[width=0.23\textwidth]{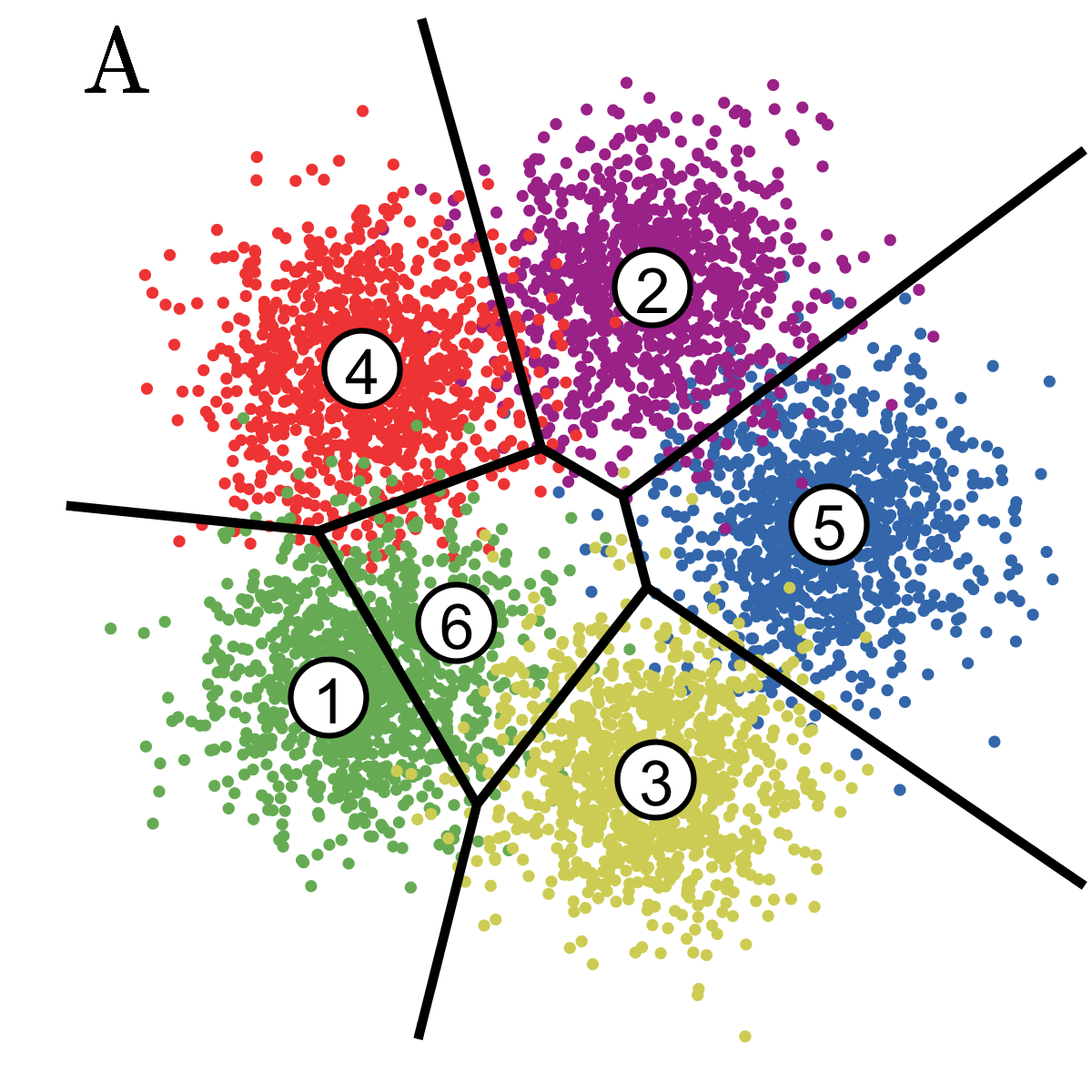}
\includegraphics[width=0.23\textwidth]{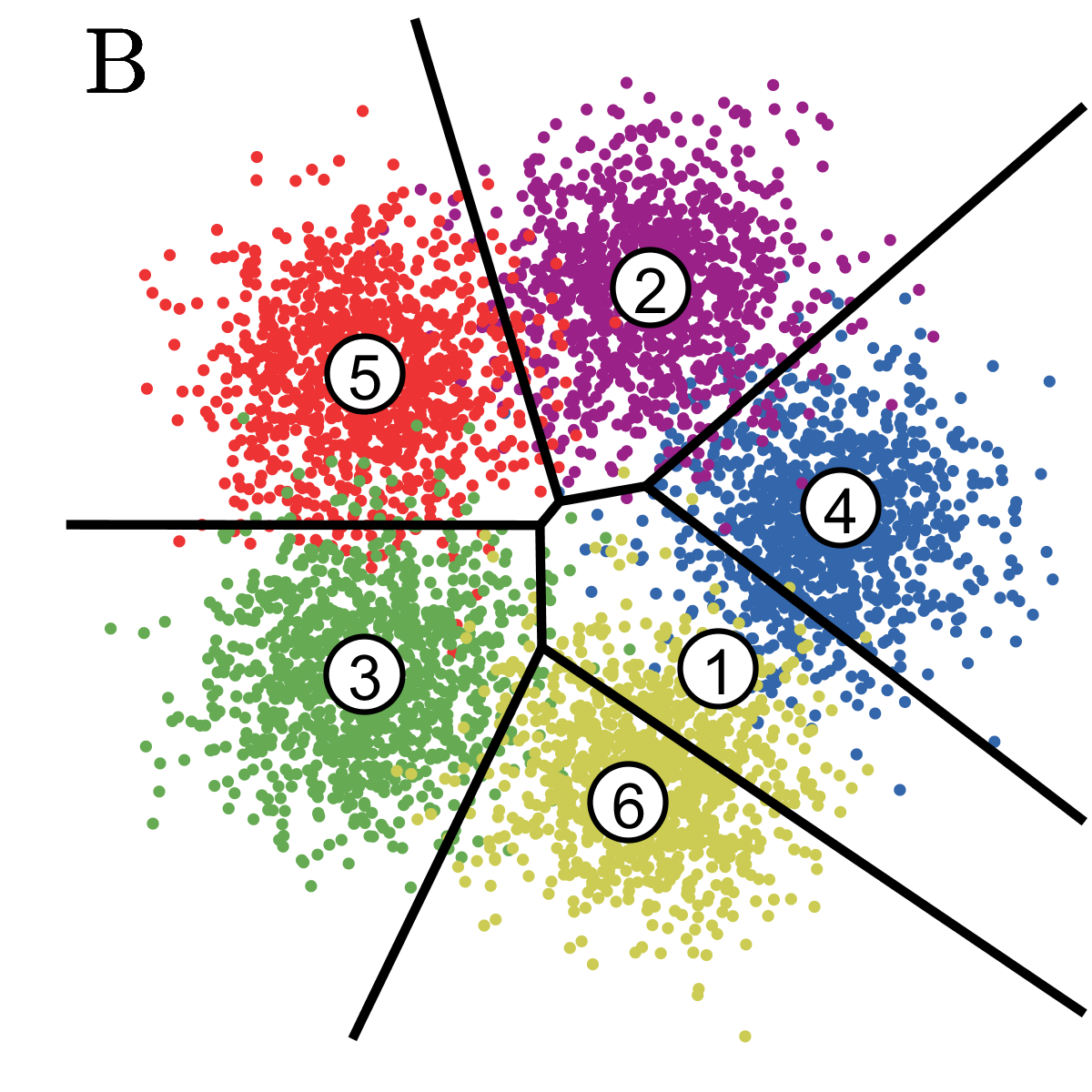}
\caption{Examples of \texttt{k-means} clustering at $k = 6 > k_{true}$. The algorithm has split the green true cluster in the example on the left, and the yellow true cluster in the example on the right. The $k$-means centres are marked by filled white circles. The boundaries between \texttt{k-means} clusters are marked by straight black lines.}
\label{fig:sim6}
\end{figure}

Figure \ref{fig:sim6} shows two examples of the solutions found at $k = 6 > k_{true}$. The algorithm has split a true cluster in both cases: green in solution A on the left, and yellow in solution B on the right. The splits appear to have a lesser effect on neighbouring \texttt{k-means} clusters than the mergers at $k = 4$, in that there is less contamination overall. Contingency table \ref{tab:sim6} reveals that the solutions are more strongly associated with one another than those solutions found at $k = 4$, as the split in one solution fits more cleanly into a whole cluster in the other. For these two $k = 6$ solutions (A and B), we calculate $\chi^{2} = 18017.03$ and (with $N = 5000, k = 6$) $V = 0.85$.

\begin{table}
\caption{Contingency table, comparing solutions A and B generated at $k = 5$ and shown in figure \ref{fig:sim6}.}
\label{tab:sim6}
\centering
\begin{tabular}{l | r r r r r r}
\hline
Cluster		& B1$_{6}$	& B2$_{6}$	& B3$_{6}$	& B4$_{6}$	& B5$_{6}$	& B6$_{6}$	\\
\hline
A1$_{6}$	& $0$		& $0$		& $658$		& $0$		& $0$		& $8$		\\
A2$_{6}$	& $1$		& $988$		& $0$		& $0$		& $0$		& $0$		\\
A3$_{6}$	& $258$		& $0$		& $0$		& $0$		& $0$		& $658$		\\
A4$_{6}$	& $0$		& $1$		& $0$		& $0$		& $954$		& $0$		\\
A5$_{6}$	& $148$	 	& $3$		& $0$		& $844$		& $0$		& $0$		\\
A6$_{6}$	& $34$	 	& $4$		& $338$		& $0$		& $37$		& $31$		\\
\hline
\end{tabular}
\end{table}

\begin{figure}
\centering
\includegraphics[width=0.45\textwidth]{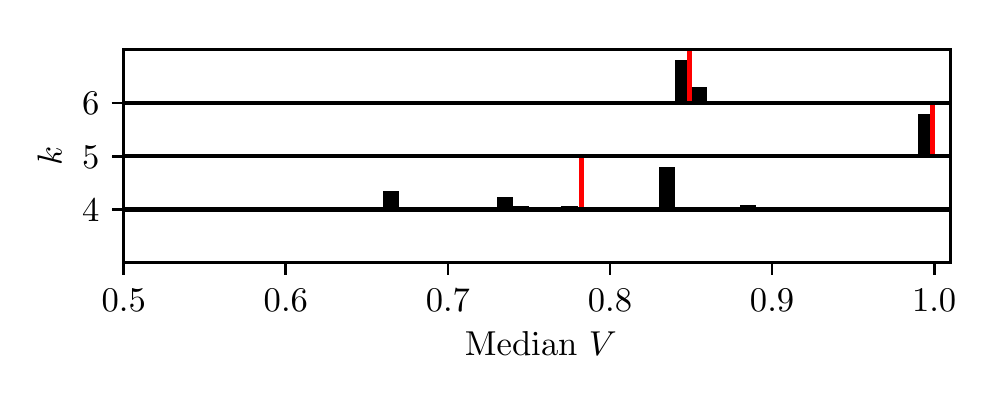}
\caption{Stability map of \texttt{k-means} clustering for the simulated data set at $k = 4, 5,$ and $6$. We calculate the median $V$ of each solution with respect to all other solutions at the same $k$. The distributions of all $200$ medians at each $k$ are represented using histograms plotted along each of the horizontal black baselines. The heights of the histograms are normalised. Additionally, we show the means of these distributions as vertical red lines. The solution at $k = 5$ stands out as being particularly stable, indicative of the true structure of the simulation.}
\label{fig:simstab}
\end{figure}

We summarise these results using a stability map (figure \ref{fig:simstab}). We reemphasise that the key element of this plot for distinguishing stable and unstable values of $k$ is the gap across all distributions in median $V$. The distribution of solutions at $k = 5$, showing that all $200$ initialisations converged to the same stable solution (within $4$ points), is clearly indicative of the true structure of the simulation. The distributions of solutions at $k = 4$ and $k = 6$, indicating that they are unstable, reflect that there is no objectively correct way to divide the five true clusters into four or six given the symmetry of the simulation. The distribution at $k = 6$ is narrower because splits affect the accuracy of the other \texttt{k-means} clusters less than mergers. With the benefit of knowing the true structure of the simulation, we could remerge the splits at $k = 6$ and achieve a better approximation to the $k = 5$ solutions than if we were to split merges at $k = 4$. For more complicated samples, involving more features, this effect would be more difficult to discover and exploit.

Given that the same solution may arise several times over a large number of initialisations, one may opt to select the modal solution as the most optimal instead of that with the lowest $\phi$. In practice, we find that one criterion implies the other; the most compact clusters tend to emerge most often anyway (thanks to our choice of initialisation technique). We retain $\phi$ as our criterion for optimal clustering at given values of $k$.

\section{Bootstrap experiment}
\label{app:bs}

\begin{figure}
\centering
\includegraphics[width=0.45\textwidth]{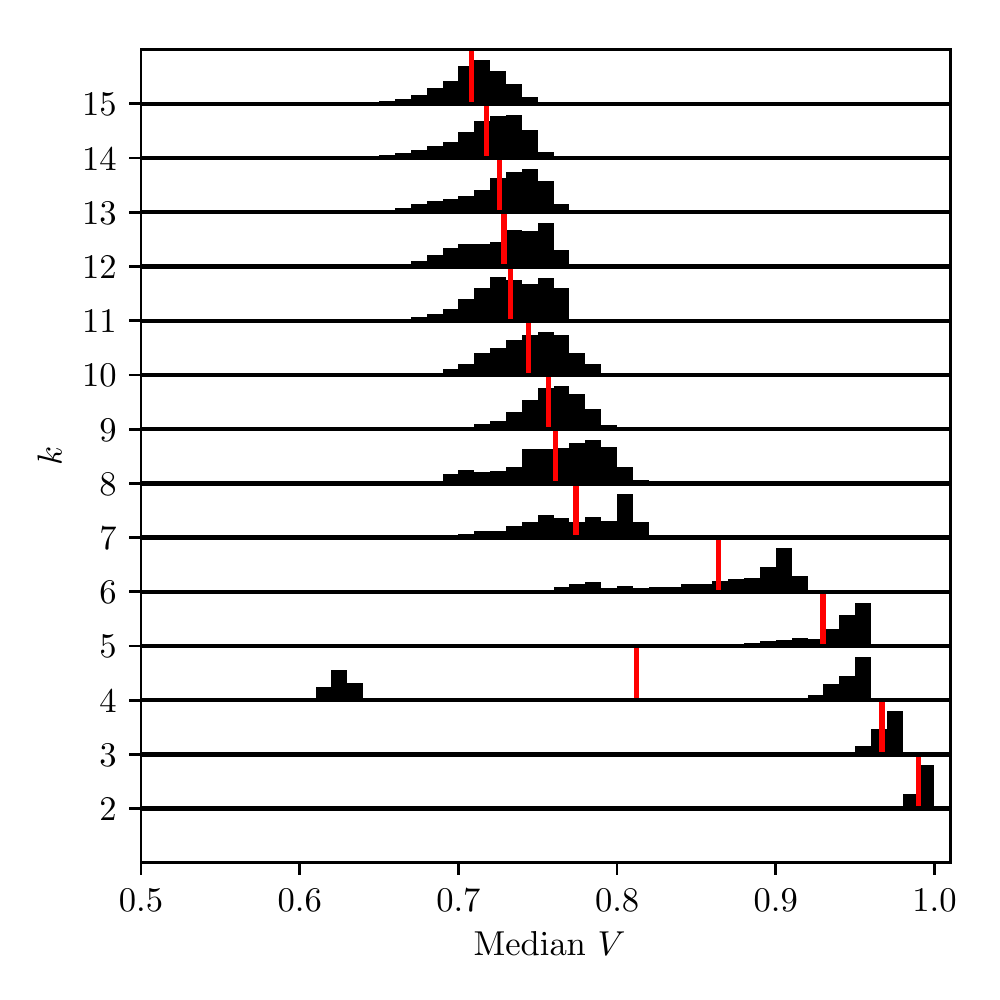}
\caption{Stability map of \texttt{k-means} clustering for our bootstrapped sample at $k = 2$ through $k = 15$. We calculate the median $V$ of each solution with respect to all other solutions at the same $k$. The distributions of all $7338$ medians at each $k$ are represented using histograms plotted along each of the horizontal black baselines. The heights of the histograms are normalised. Additionally, we show the means of these distributions as vertical red lines. Solutions at $k = 2,3,5,$ and $6$ remain most stable following application of the bootstrap method to our sample.}
\label{fig:bsstab}
\end{figure}

In order to estimate the uncertainties on the centroids reported in table \ref{tab:c}, we apply the bootstrap method to our sample of galaxies. The method resamples our original sample with replacement, such that the same galaxy may be selected more than once. We select $7338$ observations in this manner. We run \texttt{k-means} once on this new sample, retaining the centroids, and then partition our original sample according to these centroids. This whole process is itself repeated $7338$ times. The stability map for the bootstrap experiment is shown in figure \ref{fig:bsstab}.

The distributions of solutions at all values of $k$ are shifted to lower stabilities following application of the bootstrap method. This is in comparison with the distributions generated purely from the original sample, shown in figure \ref{fig:stab}. Solutions at $k = 2, 3, 5,$ and $6$ remain the most stable in figure \ref{fig:bsstab}, though solutions at $k = 6$ exhibit a more significant reduction in stability than solutions at $k = 2, 3,$ and $5$ due to the increased local dependency of \texttt{k-means} with a higher numbers of centres. Solutions at $k = 4$ retain their bimodal structure in stability. The distribution of solutions at $k = 7$, which exhibited a stable component in figure \ref{fig:stab}, is now uniformly unstable to the same extent as the distributions at higher values of $k$, justifying our decision to exclude solutions from $k = 7$ from our analyses in section \ref{sec:res}.

For the lower and upper uncertainties in table \ref{tab:c}, we calculate the $16$th and $84$th percentiles respectively of the $7338$ centroids, in each of the five features. From these, we subtract the original centroids we calculate.

\section{Postage Stamps}
\label{app:ps}

Here we present example postage stamps of galaxies in each of the clusters in each of the solutions. The three-colour stamps are made using $r$- and $g$-band imaging from the Kilo-Degree Survey \citep{DEJONG+2013}, and a mean of the two bands as the central colour channel. The stamps enclose each galaxy to $2.5$ times its Kron radius. The examples we choose are those that are best represented by the cluster centroids; they are nearest to the centroids in our five dimensional feature space.

\begin{figure*}
\centering
\includegraphics[width=0.9\textwidth]{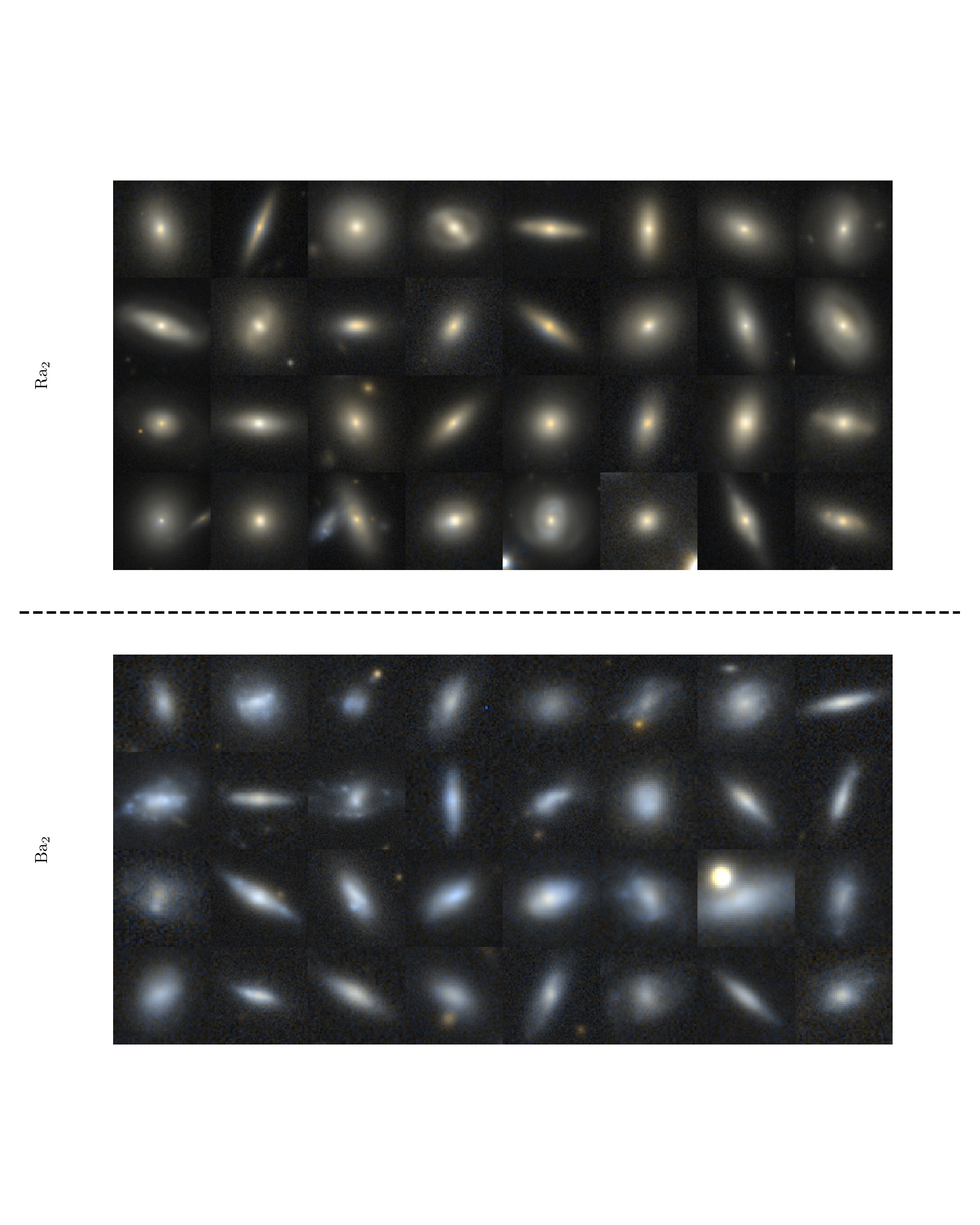}
\caption{Example postage stamps of galaxies in each of the clusters in $k = 2$. The dashed black line separates the two superclusters that \texttt{k-means} finds. See section \ref{sec:k2} for discussion.}
\label{fig:ps2}
\end{figure*}

\begin{figure*}
\centering
\includegraphics[width=0.9\textwidth]{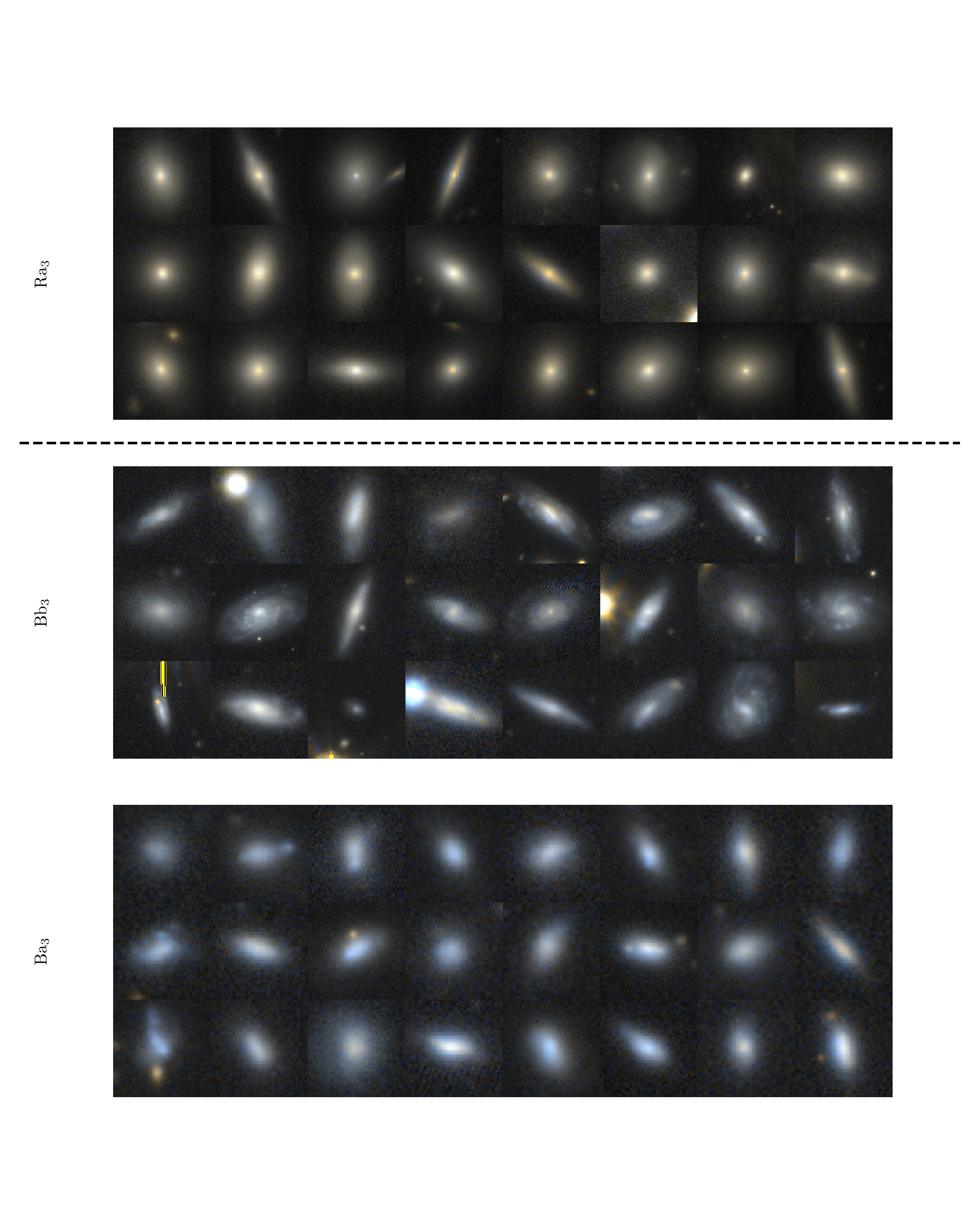}
\caption{Example postage stamps of galaxies in each of the clusters in $k = 3$. The dashed black line separates the two superclusters that \texttt{k-means} finds. See section \ref{sec:k3} for discussion.}
\label{fig:ps3}
\end{figure*}

\begin{figure*}
\centering
\includegraphics[width=0.9\textwidth]{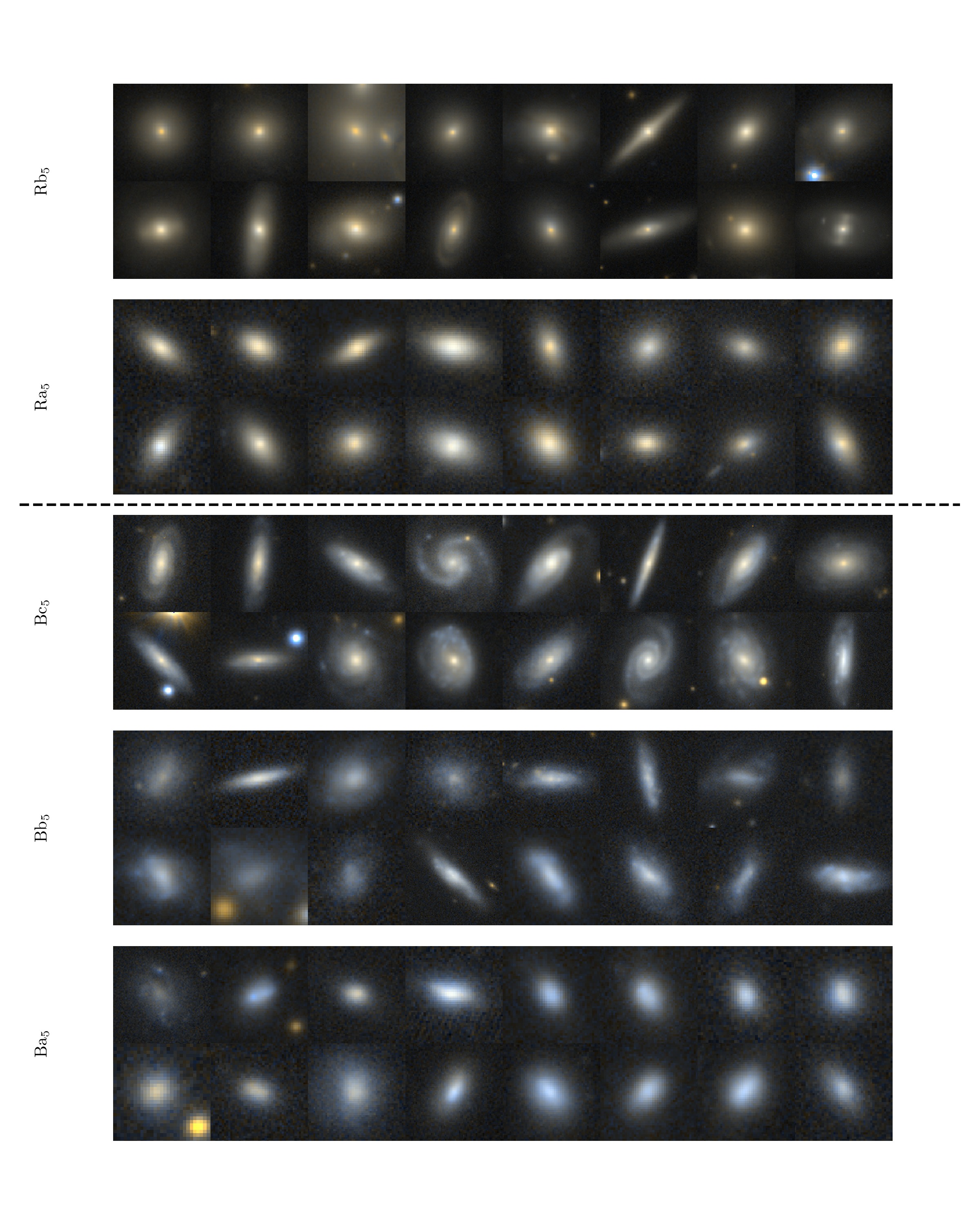}
\caption{Example postage stamps of galaxies in each of the clusters in $k = 5$. The dashed black line separates the two superclusters that \texttt{k-means} finds. See section \ref{sec:k5} for discussion.}
\label{fig:ps5}
\end{figure*}

\begin{figure*}
\centering
\includegraphics[width=0.9\textwidth]{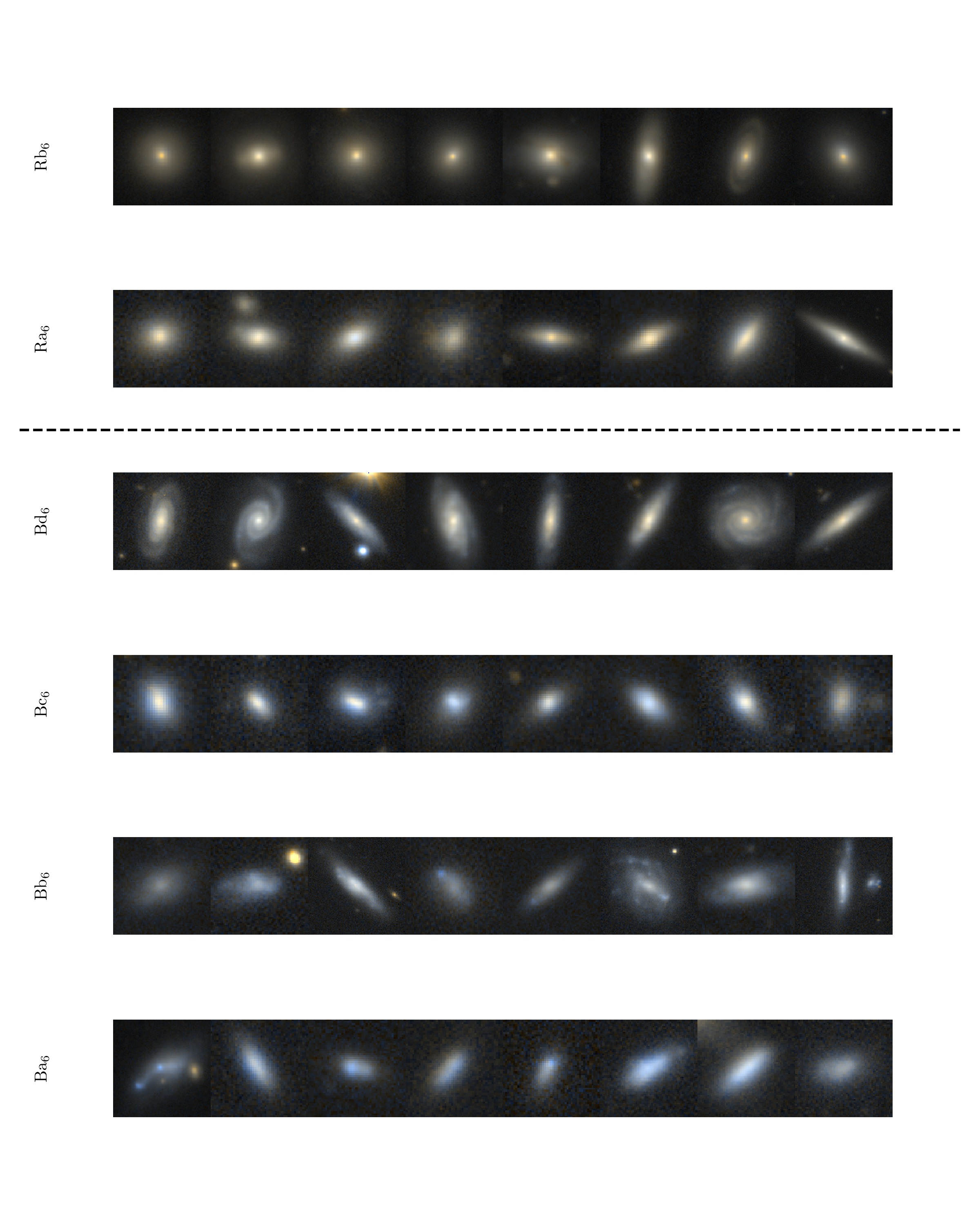}
\caption{Example postage stamps of galaxies in each of the clusters in $k = 6$. The dashed black line separates the two superclusters that \texttt{k-means} finds. See section \ref{sec:k6} for discussion.}
\label{fig:ps6}
\end{figure*}

\bsp	
\label{lastpage}
\end{document}